\documentclass[aps,prd,twocolumn,superscriptaddress,showpacs,nofootinbib,eqsecnum,amsfonts,amsmath]{revtex4-1}

\usepackage{epsfig}

\usepackage{amsfonts}
\usepackage{amsmath}
\usepackage{amssymb}
\usepackage{slashed}
\usepackage{graphicx,color}
\usepackage{float}
\usepackage{hyperref}

\def\nn{\nonumber}
\def\be{\begin{equation}}
\def\ee{\end{equation}}
\def\bea{\begin{eqnarray}}
\def\eea{\end{eqnarray}}
\def\ms{\overline {\rm MS}}

\begin{document}

\title{ Asymptotically Free Theory with Scale Invariant Thermodynamics  } 

\author{Gabriel N. Ferrari}
\affiliation{Departamento de F\'{\i}sica, Universidade Federal de Santa
  Catarina, 88040-900 Florian\'{o}polis, Santa Catarina, Brazil}

\author{Jean-Lo\"{\i}c Kneur}
\affiliation{Laboratoire Charles Coulomb (L2C), UMR 5221 CNRS-Universit\'{e} Montpellier, 34095 Montpellier, France}

\author{Marcus Benghi Pinto}
\affiliation{Departamento de F\'{\i}sica, Universidade Federal de Santa
  Catarina, 88040-900 Florian\'{o}polis, Santa Catarina, Brazil}
  
\author{Rudnei O. Ramos} 
\affiliation{Departamento de F\'{\i}sica Te\'orica, Universidade do
  Estado do Rio de Janeiro, 20550-013 Rio de Janeiro, RJ, Brazil}
\affiliation{Physics Department, McGill University, Montreal, QC, H3A 2T8, Canada}

\begin{abstract}

 A recently developed variational resummation technique, incorporating
 renormalization group properties consistently, has
 been shown to solve the scale
 dependence problem that plagues the evaluation of  thermodynamical
 quantities, e.g., within the framework of approximations such as in
 the hard-thermal-loop resummed perturbation theory.  This
 method is used in the present work to evaluate  
   thermodynamical quantities within the  two-dimensional nonlinear
 sigma model, which, apart from providing a technically simpler
 testing ground, shares some common features  with Yang-Mills
 theories, like asymptotic freedom, trace anomaly  and the nonperturbative generation
 of a mass gap.  The present application confirms that nonperturbative
 results can  be readily generated solely by considering  the 
   lowest-order (quasi-particle) contribution to the thermodynamic
 effective potential, when this quantity is required to be
 renormalization group invariant. We also show that when the
 next-to-leading correction from the method is accounted for, the
 results  indicate convergence, apart from  optimally preserving,
   within the approximations here considered, the sought-after scale
 invariance. 

\end{abstract}

\maketitle

\section{Introduction}

The theoretical description of the  quark-gluon plasma phase
transition requires the use of nonperturbative methods, since the use
of perturbation theory (PT) near the transition   is unreliable. Indeed,
it has been observed that when successive terms in the weak-coupling
expansion are added, the  predictions  for the  pressure  fluctuate
wildly  and  the  sensitivity  to  the renormalization  scale, $M$,
grows (see, e.g., Ref.~\cite{Trev} for a review).  Due to the
asymptotic freedom phenomenon PT only produces convergent results   at
temperatures many orders of magnitude larger than the critical
temperature for deconfinement.  At the same time, the development of
powerful computers and numerical techniques offers the possibility to
solve  nonperturbative problems {\it in silico} by discretization of
the  space-time onto a lattice  and then performing numerical
simulations employing the methods of lattice quantum chromodynamics
(LQCD). 

So far, LQCD  has been very successful in the description of  phase
transitions at finite temperatures and near vanishing baryonic
densities, generating results~\cite{aoki} that can be directly used
for interpreting the experimental outputs from  heavy ion collision
experiments, envisaged to  scan over this particular region of the
phase diagram. However, currently, the complete description of
compressed baryonic matter cannot be achieved due to the so-called
sign problem~\cite{signpb},  which is an unfortunate situation,
especially in view of the new experiments,  such as the Beam Energy
Scan  program at the Relativistic Heavy-Ion Collider facility.  In
this case, an alternative is to use approximate but more analytical nonperturbative
approaches. One of these is to  reorganize the series using a
variational approximation,  where  the result of a related solvable
case is rewritten in terms of a variational parameter, which in
general  has no intrinsic physical meaning and can be viewed as a
Lagrangian multiplier that allows for optimal (nonperturbative)
results to be obtained. 

In the past  decades nonperturbative methods based on related
variational methods  have been employed under different names, such as
the linear delta expansion (LDE)~\cite{lde},  the optimized
perturbation theory (OPT)~\cite{opt,pms}, and the screened
perturbation theory (SPT)~\cite{SPT,SPT3l}.  The application of these
methods starts by using a peculiar interpolation of the original
model.  {}For instance, taking the $\lambda \phi^4$  scalar theory as
an example, the basic idea is to add a Gaussian term $(1-\delta) m^2
\phi^2$ to the potential energy density, while rescaling the coupling
parameter  as $\lambda \to \delta \lambda$. One then treats the terms
proportional to $\delta$ as interactions,  using $\delta$ as a
bookkeeping parameter to perform a series expansion around the exactly
solvable theory represented by the ``free" term, $m^2 \phi^2$. At the
end, the bookkeeping parameter $\delta$  is set to its original value
$(\delta=1)$, while optimally fixing the dependence upon the arbitrary
mass $m$ (that remains at any finite order in such a modified
expansion) by an appropriate variational criterion. The idea is to explore the easiness of
perturbative evaluations (including renormalization) to get higher
order contributions that usually go beyond the topologies considered
by traditional nonperturbative\footnote{In the present context in the following by 'nonperturbative' 
we mean its specific acceptance as a method 
giving a non-polynomial dependence in the coupling well beyond standard perturbative expansion, like 
the $1/N$ expansion typically.}
techniques, such as the large-$N$
approximation. This technique
has been used to describe successfully a variety of different physical
situations, involving phase transitions in a variety of different
physical systems, such as in the determination of the critical
temperature for homogeneous Bose gases~\cite{beccrit,bec2},
determining the critical dopant concentration in
polyacetylene~\cite{poly}, obtaining the phase diagram  of magnetized
planar fermionic systems~\cite{GNmag}, in the analysis of phase
transitions in general~\cite{PTopt}, in the evaluation of quark
susceptibilities within effective QCD inspired models~\cite{tulio}, as
well as in other applications related to effective models for
QCD~\cite{Kneur:2010yv}.  Of course, due to gauge invariance issues, one
cannot simply consider a  gluonic local mass when applying SPT or OPT
to QCD. Nevertheless this procedure can be done in a
gauge-invariant manner by applying it on the previously well-defined
gauge-invariant framework of Hard Thermal Loop (HTL)\cite{HTLorig}, and it resummation, HTLpt, 
was developed over one decade
ago~\cite{htlpt1,HTLPT3loop}.  Recently, this approximation has
been evaluated up to three-loop order in the case of hot and dense
quark matter~\cite{HTLPTMU}, giving results in reasonable agreement with LQCD
for the pressure and other thermodynamical quantities. 
The SPT method has even been pushed to four-loop order in the scalar $\phi^4$ model~\cite{SPT4l}. 
However, the results
of resummed HTLpt exhibit a strong sensitivity to the
arbitrary renormalization scale $M$ used in the regularization
procedure. This is highly desirable to be reduced if one wants to
convert these available high order HTLpt results into
much more precise and reliable nonperturbative ones and, likewise, to
be  consistent with expected renormalization group invariance
properties. One could hope that the situation would improve by considering higher
order contributions, but exactly the opposite has been observed to
occur. As recently illustrated in
Refs.~\cite{htlpt1,HTLPT3loop,HTLPTMU}, at three loop order HTLpt predicts results close to LQCD simulations
for moderate $T\gtrsim 2T_c$ at the ``central" energy scale value $M=2\pi
T$, such that large logarithmic terms are minimized, but this nice
agreement is quickly spoiled when varying the scale even by a rather
moderate amount.  

A solution to this problem has been recently
proposed, by generalizing to thermal theories a related variational resummation approach, 
Renormalization Group Optimized Perturbation Theory (RGOPT). Essentially the novelty is that it
restores perturbative scale invariance at all stages of the calculation, in particular 
when fixing the arbitrary mass parameter from the variational  procedure described above, 
where it is induced by solving the (mass) optimization prescription consistently with the
renormalization group equation. 
The RGOPT was first developed at vanishing temperatures
and densities in the framework of the  Gross-Neveu (GN)
model~\cite{JLGN}, then within QCD to estimate the basic scale ($\Lambda_{\overline {\rm
    MS}}$\cite{JLQCD1}, or equivalently the QCD coupling $\alpha_S$). 
At three-loop order it gives accurate results~\cite{JLQCD2}, compatible
with the $\alpha_S$ world averages. The method has also given a precise evaluation of the quark
condensate~\cite{JLqq}.  More recently some of the present
authors have shown, in the context of the $\lambda \phi^4$ scalar
model, that the RGOPT is also compatible with the introduction of
control parameters such as the temperature~\cite{jlprl,jlprd}. 
The RGOPT and SPT predictions for the pressure
have been compared, showing how the RGOPT indeed drastically improves
over the generic scale dependence problem of thermal perturbation
theories at increasing perturbative orders. 

{}We also remark that within more standard variational approaches
such as OPT, SPT and HTLpt, the optimization process can allow for
multiple solutions, including complex-valued ones, as one considers
higher and higher orders. Accordingly, in some cases, one is forced to
obtain optimal results by using an alternative criterion, for example,
by replacing the variational mass with a purely perturbative screened
mass~\cite{SPT3l,HTLPT3loop,HTLPTMU}, but at the expenses
of  potentially loosing valuable nonperturbative information.  As
shown in Ref.~\cite{JLGN,JLQCD2}, the RGOPT may also avoid  this serious
problem,  by requiring asymptotic matching of the optimization solutions with the
standard perturbative behavior for small couplings.

Various approaches have been made earlier to improve the higher order stability and scale dependence
of thermal perturbation theories. For instance the nonperturbative RG (NPRG) framework (see e.g. ~\cite{NPRG}
for the $\phi^4$ model) should in principle give
exactly scale invariant results by construction, if it could be performed exactly. 
 But solving the relevant NPRG equations for thermal QCD beyond approximative truncation schemes
appears very involved.
Other more perturbative attempts have been made to improve the perturbative scale dependence of 
thermodynamical QCD quantities, not necessarily relying on a variational or HTLpt resummation framework: rather 
essentially using RG properties of standard thermal perturbation theories (see, e.g., ~\cite{RG_E0QCD,RArriola}).
Our approach also basically starts from standard perturbative expressions, and perturbative RG properties 
(which is one advantage since many already available higher order thermal perturbative results can be exploited). 
But it is very different from the latter approaches, 
due to the crucial role of the variational (optimization) procedure, rooted within a massive scheme.
An additional bonus provided by our procedure, as we will illustrate here, 
is that some characteristic nonperturbative features are already
provided at the lowest (``free gas") order.

In this work we apply the RGOPT to the nonlinear sigma model (NLSM)
in 1+1 dimensions at finite temperatures in order to pave the way for
future applications concerning other asymptotically free theories,
such as thermal QCD. Apart from asymptotic freedom, the NLSM and QCD have
other similarities, like the generation of a mass gap and trace
anomaly. In the previous RGOPT finite temperature
application~\cite{jlprd,jlprl},  the numerical results for the
pressure were mainly expressed as functions of the coupling, as done
in the usual SPT applications to scalar theories. Here, on the other
hand, we  perform an investigation more reminiscent of typical HTLpt
applications to hot QCD (see, e.g.,
Refs.~\cite{htlpt1,HTLPT3loop,HTLPTMU}), by  mainly concentrating on
the (thermodynamically) more appealing $P-T$ plane.  Another, more
technical but welcome feature of considering the NLSM is that, up to
two-loop order, the relevant thermal integrals are simple and exactly
known (at least for the pressure and derived quantities),  which
allows for a rather straightforward study of the full temperature
range with our method.  As a simple model that has been studied many
times before in the context of its critical properties and
renormalization group results, the NLSM makes then a perfect test
ground for benchmarking the RGOPT when compared to other
nonperturbative methods. 

It is worth mentioning that for HTLpt applied to QCD  at two-loop order and
beyond, results~\cite{HTLPT3loop} are only available in the high-$T$
approximation regime, not to mention the rather involved
gauge-invariance framework required by the method.  At three-loop
order the NLSM starts to involve more complicated integrals, but this
is beyond the present scope, and two-loop order RGOPT, that we will 
carry out in the present work, will be enough to illustrate the RGOPT efficiency.
Although, to the best of our knowledge, SPT (or its high-$T$ expansion
variant more similar to HTLpt) has not been applied previously in the
NLSM framework, we found it worth to derive and compare in some detail
such SPT/HTLpt results with the RGOPT results in the present model.
This is useful in order to emphasize the improvements of RGOPT that
are generic enough to be appreciated in view of QCD applications.  In
this work, we also investigate how the RGOPT performs with respect to
other thermodynamical  characteristics, like the Stefan-Boltzmann
limit and the trace anomaly, among others, which were not investigated
in Refs.~\cite {jlprd,jlprl}.  

As we will illustrate, the scale
invariant results obtained in the present application give further
support to the method as a robust analytical nonperturbative
approach to thermal theories.  Bearing in mind that
the RGOPT is rather recent, we will also perform the basic
derivation in a way that the present work may also serve as a
practical guide for further applications in other thermal field
theories. 

This paper is organized as follows. In Sec.~\ref{sec2} we briefly
review the NLSM. Then, in Sec.~\ref{sec3},  we perform the
perturbative evaluation of the pressure to the first nontrivial order
and discuss the perturbative scale invariant construction. In
Sec.~\ref{sec4} we modify the perturbative series in order to make it
compatible with the RGOPT requirements. The optimization procedure is
carried out in Sec.~\ref{sec5} for arbitrary $N$ up to two-loop order,  where we also derive the large-$N$
approximation, the  standard perturbation (PT), and the SPT/HTLpt
alternative approaches, also up to two-loop order for comparison. Our numerical results are
presented and discussed in Sec.~\ref{sec6}, where we compare the
previous different approximations as well as the next-to-leading (NLO) order of the
$1/N$-expansion~\cite{warringa}, for $N=4$, which is a physically appealing choice 
beyond the more traditional $N=3$ continuum limit of the Heisenberg model.
Then we also compare our RGOPT results 
with lattice simulation results, apparently only available for $N=3$~\cite{giacosa}.  {}Finally, in Sec.~\ref{sec7},
we present our conclusions and final remarks.

\section{The NLSM in 1+1-dimensions}
\label{sec2}

The two-dimensional
NLSM partition function can be written as~\cite{nlsmrenorm,ZJ}

\begin{equation}
Z \!=\! \int \! \prod_{i=1}^{N} {\cal D} \Phi_i(x) \exp \left [  \frac
  {1}{2 g_0} \!\!\int \! d^2 x (\partial \Phi_i)^2 \right ] \delta
(\sum_{i=1}^N \Phi_i \Phi_i -1),
\label{action}
\end{equation}
where $g_0$ is a (dimensionless) coupling and the scalar field is parametrized as
$\Phi_i=(\sigma,\pi_1,...,\pi_{N-1})$. In two-dimensions the theory is
 renormalizable~\cite{nlsmrenorm} and also, according to the Mermin-Wagner-Coleman
 theorem~\cite{mermin,coleman},  no  spontaneous symmetry breaking of
 the global $O(N)$ symmetry can take place (at any coupling value). The action is invariant under
$O(N)$ but using the constraint, $\sigma(x)= (1-  \pi_i^2)^{1/2}$, 
 in order to define the perturbative expansion, 
breaks the symmetry down to
$O(N-1)$. This is accordingly an artifact of perturbation theory, and truly nonperturbative
quantities, when calculable, should exhibit the actually unbroken $O(N)$ symmetry~\cite{ZJ}, 
as shown by the
nonperturbative exact mass gap at zero temperature~\cite{mgapTBA}.
Thus the perturbation theory describes at first $N-1$ Goldstone bosons,  and one may introduce,
for later convenience, an infrared regulator, $m_0^2$, coupled to
$\sigma$. In this case the partition function becomes 

\begin{equation}
Z(m) = \int d\pi_i(x)  \left [1- \pi_i^2(x) \right ]^{-1/2}  \exp
[-{\cal S}(\pi,m)],
\end{equation}
where the (Euclidean) action is ${\cal S}(\pi,m) = \int d^2 x {\cal
  L}_0$ and, upon rescaling $\pi_i \to\sqrt{ g_0} \pi_i$, the  bare
Lagrangian density is 

\begin{equation}
{\cal L}_0 = \frac{1}{2} (\partial \pi_i)^2 + \frac { g_0 (\pi_i
  \partial \pi_i)^2}{2(1- g_0 \pi_i^2)} - \frac{m_0^2}{g_0}
  \left (1- g_0 \pi_i^2 \right)^{1/2}  .
\label{NLSMLag}
  \end{equation}
The above Lagrangian density can
be expanded to order-$g_0$ yielding

\begin{equation}
{\cal L}_0 \!=\! \frac{1}{2}  \left[ (\partial \pi_i)^2 + m_0^2
  \pi_i^2 \right] + \frac {g_0 m_0^2}{8} (\pi_i^2)^2 + \frac{g_0}{2 }
(\pi_i \partial \pi_i)^2 -{\cal E}_0 ,
\end{equation}
 where for later notational convenience we designate as ${\cal E}_0\equiv m^2_0/g_0$ 
the field-independent term, originating at lowest order from expanding the square root in Eq.~(\ref{NLSMLag}).
 At first, one may 
think that such field-independent ``zero-point" energy term could be dropped innocuously (as is indeed sometimes  
assumed in the literature~\cite{makino}). However, as we will examine  below it is important to keep this term 
since it plays a crucial role for consistent perturbative RG properties. 

In Euclidean spacetime the {}Feynman rules of the model can be found,
e.g., in refs~\cite{ZJ,PSbook}.  The Euclidean four-momentum, in the
finite temperature Matsubara's formalism~\cite{kapusta}, is $p_{0,Eucl}
\equiv  \omega_n$,  where $\omega_n=2\pi n T$ are the bosonic
Matsubara frequencies $(n=0,\pm 1, \pm 2\cdots)$   and $T$ is the
temperature.  In this work,  the divergent integrals are regularized
using dimensional regularization (within the minimal subtraction
scheme $\overline {\rm MS}$), which at finite temperature and $d=2-
\epsilon$ dimensions, can be implemented by using

\begin{equation}
\int \frac {d^2 p} {(2\pi)^2} \to T \,
\hbox{$\sum$}\!\!\!\!\!\!\!\int_{\bf p} \equiv T\left ( \frac
     {e^{\gamma_E} M^2}{4\pi} \right)^{\epsilon/2}
     \sum_{n=-\infty}^{+\infty} \int \frac {d^{1-\epsilon} 
         p}{(2\pi)^{1-\epsilon}},
\end{equation}
 where $\gamma_E$ is the Euler-Mascheroni constant and $M$ is the
 $\overline {\rm MS}$ arbitrary regularization energy scale.  At
 finite temperatures this model has been first studied by Dine and
 Fischler~\cite{dine} in the context of the PT and also in the large $N$
 approximation.

\section{ Perturbative Pressure and Scale Invariance}\label{sec3}
 \begin{figure}[htb!]
\includegraphics[scale=0.5]{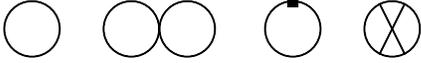}
\caption[long]{\label{fig1} Feynman diagrams contributing to the
  perturbative pressure at ${\cal O}(g)$. The first term represents $P_0(m_0)$, the second, $P_1(m_0, g_0)$, 
  the third term represents the self-energy counterterm $P_0^{\rm CT}$ (obtained 
  from expanding $Z_m$ to first order in $P_0(m_0=Z_m\,m)$), while the fourth term
  represents the zero point contribution ${\cal E}_0(g_0)$ to
  Eq.~(\ref {press2l}). }
\end{figure}
Considering the  contributions displayed in
{}Fig.~\ref{fig1}, one can write the pressure up to order ${\cal O}(g_0)$ as

\begin{equation}
 P= P_0(m_0) + P_1(m_0, g_0) + {\cal E}_0(m_0, g_0) + {\cal O}(g_0^2), 
\label {press2l}
\end{equation}
where the (one-loop) zeroth-order term represents the usual free gas
type of term and it is given by

\begin{equation}
P_0(m_0) = - \frac{(N-1)}{2} I_0(m_0,T),
\label{P0}
\end{equation}
where

\begin{eqnarray}
I_0(m_0,T) =  T \, \hbox{$\sum$}\!\!\!\!\!\!\!\int_{\bf p} \ln \left(
\omega_n^2+\omega_{\bf p}^2 \right),
\end{eqnarray}
with the dispersion relation,  $\omega_{\bf p}^2 ={\bf p}^2 +m_0^2$.

At two-loop order the pressure receives the contribution from the
${\cal O}(g_0)$ term

\begin{equation}
P_1(m_0, g_0)= -(N-1) \frac{(N-3)}{8} m_0^2 g_0 I_1(m_0,T)^2 ,
\label{p1}
\end{equation}
where $I_1(m_0,T)=\partial I_0(m_0,T)/\partial m_0^2$,  as well as
from the counterterm insertion contributions in the one-loop pressure. To this
perturbative order, one has $g_0=Z_g g \equiv g$  and thus just
a mass counterterm insertion contribution in the one-loop pressure to deal with. It leads to a
counterterm $P^{\rm CT}_0$ that can  be readily obtained by replacing $m_0 =
Z_m m$ in $P_0$ and expanding it up to first-order, $P_0(m_0=Z_m\,m)= P_0(m)+P^{\rm CT}_0(m,g)$, where explicitly

\begin{equation}
P^{\rm CT}_0(m, g) = \frac{(N-1)(N-3)}{8\pi \epsilon} m^2 g I_1(m,T) ,
\label{pct}
\end{equation}
upon using~\cite{hikami,ZJ} (our convention is $d=2-\epsilon$) 
\begin{equation}
Z_m= 1 - \frac{g}{8\pi } (N-3)\frac{1}{ \epsilon} + {\cal O}(g^2).
\label{Zm}
\end{equation}

Then, when performing the sum over the Matsubara's frequencies within
the ${\overline {\rm MS}}$ scheme one obtains for the loop momentum
integrals $I_0$ and $I_1$ appearing in the above expressions, the
explicit results

\begin{eqnarray}
& & I_0(m_0,T) =  \frac{ m_0^2}{2\pi} \left \{ \frac{1}{\epsilon} + 
  \frac{1}{2} -\ln (\frac{m_0}{M}) \right. \nonumber \\
&& \left.  +\epsilon \left[ \frac{1}{4}+\frac{1}{2} \ln (\frac{m_0}{M})
  \left( \ln (\frac{m_0}{M}) -1 \right) \right]   \right \} \nn \\ 
  && + T^2 \frac{2}{\pi} J_0(m/T) ,
\end{eqnarray}
and 

\begin{eqnarray}
 I_1(m,T) &=& \frac{1}{2 \pi} \left \{ \frac{1}{\epsilon} - \ln \left (
\frac {m}{M} \right ) + \frac{\epsilon}{2}\left [ \ln^2 \left ( \frac {m}{M} \right ) + \frac{\pi^2}{24} 
   \right]  \right \} \nn
\\ &-&\frac{1}{\pi} J_1(m/T),
 \label{I1def}
\end{eqnarray}
where, in the above expressions, the thermal integrals $J_0(x)$ and
$J_1(x)$ read, respectively,
 
\begin{eqnarray}
J_0(x) = \int_0^{\infty} dz \ln \left( 1- e^{-\omega_z} \right),
\end{eqnarray}
and

\begin{eqnarray}
J_1(x) = \int_0^{\infty} dz \frac{1}{\omega_z \left( 1- e^{\omega_z}
  \right)},
\end{eqnarray}
where we have defined the dimensionless quantity $\omega_z^2 = z^2 +
x^2$, with $z=|{\bf p}|/T$ and $x=m/T$.

Putting all together in Eq.~(\ref{press2l}), inserting $I_1(m,T)$ into Eqs.~(\ref {p1}), (\ref {pct}) and
expanding to ${\cal O}(\epsilon^0)$, one may isolate the possibly
remaining divergences contributing to the pressure  (after mass and
coupling renormalization having been performed), as

\begin{eqnarray}
 P &=& - \frac{(N-1)}{2} I_0^{\rm r}(m,T) \nn \\ &-&(N-1)
 \frac{(N-3)}{8} m^2 g \left[I_1^{\rm r} (m,T)\right]^2  \nonumber \\  &-&  (N-1)
 \frac{m^2}{(4\pi) \epsilon} \left[1  - g\frac {(N-3)}{2(4\pi)
     \epsilon}   \right ] \nn \\ &+&  \frac{m^2}{g}\,Z^2_m Z^{-1}_g ,
\label{Pbare2L}
\end{eqnarray}
where we have defined the {\it finite} quantities
 
\begin{equation}  
I_0^{\rm r}(m,T)= \frac{ m^2}{2\pi} \left [ \frac{1}{2}- \ln \left (
  \frac{m}{M} \right ) \right ] + T^2 \frac{2}{\pi} J_0(m/T) ,
\label{I0r}
  \end{equation}
and

\begin{equation}
I_1^{\rm r}(m,T)=-\frac{1}{2 \pi}  \ln \left ( \frac {{m}}{M} \right )
-\frac{1}{\pi} J_1(m/T).
\label{I1r}
\end{equation}
 Then, renormalizing finally the zero-point energy ${\cal E}_0(m_0,g_0)$, last term in 
Eq.~(\ref{Pbare2L}), gives:
\bea
&&\frac{m^2}{g}\, Z^2_m Z^{-1}_g= \frac{m^2}{g}\,Z^{-1/2}_\pi \nn \\
&& = \frac{m^2}{g}\,\left(1+ \frac{(N-1)}{4\pi \epsilon}\, g -
\frac{(N-1)(N-3)}{2(4\pi)^2 \epsilon^2} g^2 +{\cal O}(g^3)\right), \nn \\
\label{E0ren}
\eea
where we used the exact NLSM relation~\cite{nlsmrenorm} $ Z^2_m Z^{-1}_g= Z^{-1/2}_\pi$ with $\pi^2_0=Z_\pi \pi^2$
and the two-loop order~\cite{hikami} $Z_\pi$ counterterm expression. Accordingly, (\ref{E0ren}) acts as a vacuum energy 
counterterm, exactly cancelling the remaining divergences in Eq.~(\ref{Pbare2L}), so that one can write
the renormalized two-loop pressure in the compact form
\begin{equation}
P = \frac{m^2}{g} - \frac{(N-1)}{2} \left[ I_0^{\rm r}(m,T) + \frac{(N-3)}{4} m^2 g
\left[I_1^{\rm r} (m,T)\right]^2 \right].
\label{RenP}
\end{equation}
Before we proceed, we should stress that those vacuum energy (pressure) renormalization features  
in the $\ms$-scheme are peculiar to the NLSM: in contrast for a general massive model the vacuum energy (equivalently pressure) is 
not expected to be renormalized {\em solely} from the mass and coupling counterterms, such that one needs
additional proper vacuum energy counterterms. Here the latter are provided for free, by 
retaining consistently the field-independent zero-point energy ${\cal E}_0(m_0, g_0)$ already present in the Lagrangian. 
Omitting this term 
would force to introduce new minimal counterterms (i.e cancelling solely the divergent terms 
shown explicitly in (\ref{E0ren})), however missing thus the {\em finite} lowest order $m^2/g$ term that remains in the 
renormalized pressure Eq.~(\ref{RenP}). Moreover, not surprisingly
the latter term is crucial to ensure perturbative RG invariance of the renormalized pressure. 
More precisely, consider the renormalization group (RG) operator,  defined by
\begin{equation}
M \frac{d}{dM} \equiv  M \frac{\partial}{\partial M} + \beta
\frac{\partial}{\partial g} - \frac{n}{2} \zeta + \gamma_m m
\frac{\partial}{\partial m} .
\label{RG}
\end{equation} 
Applying the latter to the pressure (zero-point vacuum energy) one has $n=0$,
so that one only needs to consider the $\beta$ and $\gamma_m$ functions.
At the two-loop level,
\begin{equation}
\beta = - b_0 g^2 - b_1 g^3 + {\cal O}(g^4) ,
\label{beta}
\end{equation}
and
\begin{equation}
\gamma_m = -\gamma_0 g-\gamma_1 g^2 +{\cal O}(g^3),
\label{gammam}
\end{equation}
where the RG coefficients in our normalization are~\cite{hikami}:
\begin{eqnarray}
&& b_0=(N-2)/(2\pi),
\label{b0}
\\ && b_1=(N-2)/(2\pi)^2,
\label{b1}
\\  &&\gamma_0=(N-3)/(8\pi),
\label{gamma0}
\\ &&\gamma_1=(N-2)/(8\pi^2).
\label{gamma1}
\end{eqnarray}

It is now easy to check that applying (\ref{RG}) to Eq.~(\ref{RenP}) gives 
\begin{equation}
M \frac{d P}{dM}= {\cal O}(g^2),
\label{NLSMrem}
\end{equation}
i.e. RG invariance up to higher order (three-loop here) neglected terms. 
This is a quite remarkable feature of the NLSM, that
one can trace to the nonlinear origin of the mass term in (\ref{NLSMLag})
(footprint of the decoupled $\sigma$ field, once expressed in terms of $\pi_i$ fields). 
Accordingly this contribution contains much more than mere $\pi_i$ mass terms, in particular 
the RG properties of the finite remnant $m^2/g$ piece in Eq.~(\ref{RenP}) lead to 
Eq.~(\ref{NLSMrem}). 
In contrast, for other models with linear mass terms (like in the $\phi^4$ model typically),
the naive (perturbative) vacuum energy generally badly lacks RG invariance, already at lowest order. \\
To further appreciate these features, suppose  now that we
had dropped the peculiar NLSM zero-point term ${\cal E}_0=m^2_0/g_0$ in (\ref{NLSMLag}) 
from the perturbative calculation of the 
pressure, which is exactly the situation one generally 
deals with in other (linear) models, where such terms are simply absent from the start. 
In this case the remaining divergences in Eq.~(\ref{Pbare2L}) have
to be minimally cancelled by appropriate counterterms.  
Next applying (\ref{RG}) to the resulting finite 
pressure Eq.~(\ref{RenP}) (but now missing the very first term), since $\beta$ and $\gamma_m$ are at least of
 order-$g$, one obtains 
\begin{equation}
M \frac{d P}{dM}= -(N-1) \frac{m^2}{4\pi} +{\cal O}(g),
\label{remnant}
\end{equation}
which explicitly shows the lack of perturbative scale-invariance,
the remnant term being of leading order ${\cal O}(m^2)$. 
Such remnant terms generally occur in any massive model and are 
nothing but the manifestation that the vacuum energy of a (massive)
theory has a nontrivial anomalous  dimension in general. In this case, to  
restore RG invariance one needs to add finite contributions, perturbatively determined from RG properties
(see e.g. \cite{jlprl,jlprd} in the thermal context,
or for earlier similar considerations at vanishing temperature, \cite{RGinvMS}).
Thus (once
having minimally renormalized the remaining divergences of the pressure), 
one is lead to (re)introduce  an additional finite contribution in ${\cal  E}_0$ which, upon acting with the RG
operator Eq.~(\ref{RG}), precisely compensates the remnant anomalous
dimension terms like the lowest order one in Eq.~(\ref{remnant}).
Still pretending to ignore the initially present NLSM ${\cal E}_0$ term (or when absent like in other model cases), 
one can add a finite contribution of the
generic form $m^2 f(g)/g$ (which in minimal subtraction schemes
cannot depend explicitly on the temperature, nor on the
renormalization scale $M$, since it is entirely determined from
(integrating) the RG anomalous dimension). 
{}Following \cite {jlprd,jlprl} one can write the finite zero-point
  energy contribution, ${\cal E}_0^{\rm RG}$: 
\begin{equation}
{\cal E}_0^{\rm RG} = m^2 \sum_{k \ge 0} s_k g^{k-1}  ,
\label{skdef}
\end{equation}
and determine the coefficients $s_k$  by applying (\ref{RG}) consistently order by order.
In the present NLSM, one can easily check that it uniquely fixes the relevant coefficients
up to two-loop order, $s_0, s_1$, as
\begin{equation}
s_0 = \frac{(N-1)}{4\pi (b_0-2 \gamma_0)} = 1 ,
\label{s0}
\end{equation}
and
\begin{equation}
s_1= (b_1-2\gamma_1)\frac{s_0}{2 \gamma_0}=0,
\label{s1}
\end{equation}
(which vanishes as $b_1=2\gamma_1$ in the NLSM). \\
Thus from perturbative RG considerations, Eq.~(\ref{skdef}) with (\ref{s0}), (\ref{s1}) 
reconstructs consistently the NLSM first term of (\ref{RenP}),
originally present in our original NLSM derivation above. 
While this derivation was unnecessary for the NLSM, it illustrates the procedure needed 
for an arbitrary massive model, where such finite vacuum energy terms are 
generally absent and can be reconstructed perturbatively in 
such a way. As we will see in Sec~\ref{sec4}, the presence of this finite vacuum energy piece induces a nontrivial mass gap solution
already at lowest order, in contrast with other related variational approaches.
Its presence will be crucial to obtain some essentially nonperturbative features of the model already at lowest order.  
We remark in passing that the result $s_1=0$ (equivalently the original NLSM expression (\ref{RenP})) 
is a consequence of the peculiar RG properties of the NLSM. We anticipate that this affects  the
properties of the RGOPT solution at two-loop order, as will be
examined below  in Sec.~\ref{sec5}. 
In other models those perturbative subtraction coefficients are a priori all non-vanishing, as is the
case in various other scenarios  explored so
far~\cite{JLQCD2,JLqq,jlprl,jlprd}. \\
To conclude this section, 
we stress that the vacuum energy terms as in Eq.~(\ref{skdef}), generally required in massive renormalization schemes based  on 
dimensional regularization, have been apparently ignored in many thermal field theory applications, in particular 
in the SPT and resummed HTLpt construction~\cite{htlpt1,HTLPT3loop}, essentially based on adding a (thermal) mass term. 
In contrast our construction maintains perturbative RG invariance at all levels of the calculation:
first, by considering generically for any model the required perturbative finite subtraction (\ref{skdef}) (although
already present from the start in the peculiar NLSM case, as above explained). In subsequent step, RG invariance is maintained 
(or more correctly, restored) also
within the more drastic modifications implied by the variationally optimized perturbation framework, as we examine now.

\section{RG Optimized Perturbation Theory}
\label{sec4}
 
 To implement next the RGOPT one first modify the standard perturbative
 expansion by rescaling the infrared regulator $m$ and coupling:   \be
 m\to (1-\delta)^a m,\;\;\;\; g \to \delta g,
 \label{delta}
 \ee in such a way that the Lagrangian interpolates between a free
 massive theory (for $\delta=0$) and the original massless theory
 (for $\delta=1$)~\cite{JLQCD1}.  This procedure is similar to the one
 adopted in the standard SPT/OPT~\cite{lde,opt,SPT} or HTLpt applications,
 except for the crucial difference  that within the latter methods the
 exponent is rather taken as $a=1/2$ (for scalar mass terms) or $a=1$
 (for fermion mass terms), reflecting the intuitive notion of ``adding
 and subtracting'' a mass term linearly, but without deeper
 motivations.   In contrast, as we will recall now, the exponent $a$
 in our construction is consistently and uniquely fixed from
 requiring the modified perturbation, after performing (\ref{delta}),
 to restore the  RG invariance properties,  which generally makes it
 different from the above linear values $a=1/2$  for a scalar
 term~\footnote{Nonlinear interpolations with $a\ne 1/2$ and fixed by other 
 consistency requirements had also
   been sometimes considered
   previously\cite{beccrit,bec2,kleinert}. }. 

Before we proceed, let us first remark that since the mass parameter is being optimized by using the
variational stationary mass optimization
prescription~\cite{lde,pms,opt} (as in SPT/OPT),
\begin{equation}
\frac{\partial P^{\rm RGOPT}}{\partial m} \Bigr|_{m={\bar m}} = 0 ,
\label{pms}
\end{equation}
the RG operator acquires the {\it reduced} form

\begin{equation}
\left( M \frac{\partial}{\partial M} + \beta \frac{\partial}{\partial
  g}  \right) P^{\rm RGOPT}=0 .
\label{RGr} 
\end{equation}
which is indeed consistent for a massless theory.

Then, performing the aforementioned replacements given by
Eq.~(\ref{delta}) within the pressure Eq.~(\ref{RenP}), consistently re-expanding to lowest (zeroth) order in
$\delta$, and finally taking $\delta\to 1$,  one gets
\begin{equation}
P^{\rm RGOPT}_{ 1L} =  - \frac{(N-1)}{2} I_0^{\rm r}(m,T)+
\frac{m^2}{g} (1 - 2a).
\label{Prgopt}
\end{equation}
Now to fix the exponent $a$ we require the RGOPT pressure, Eq.~(\ref
{Prgopt}), to satisfy the {\it reduced} RG relation, Eq.~(\ref
{RGr}). This {\it uniquely} fixes the exponent to 

\begin{equation}
a = \frac{\gamma_0}{ b_0} =\frac{(N-3)}{4(N-2)},
\label{exponent}
\end{equation}
where the first generic expression in terms of RG coefficients
coincides with the value found for the similar prescription applied to
the scalar $\lambda \phi^4$ theory~\cite{jlprd,jlprl} and also  to QCD
(up to trivial normalization factors).   We indeed recall that, as
discussed in Refs.~\cite {JLQCD1,JLQCD2,JLqq,jlprl,jlprd}, the exponent
$a$ is universal for a given model as it only depends on the
first-order RG coefficients, which are renormalization scheme
independent.  {}Furthermore, at zero temperature, Eq.~(\ref{exponent})
greatly improves the convergence of the procedure at higher orders:
considering only the first RG coefficients $b_0$ and the $\gamma_0$
dependence (i.e., neglecting higher RG orders and non-RG terms), it
gives the known exact nonperturbatively resummed result at the very
first order in $\delta$ and also at any successive
order~\cite{JLQCD2}. This is not the case for $a= 1/2$ (for a scalar
model), where the convergence appears very slow, if
any\footnote{Notice also that, while in many other models, the simple
  linear interpolation with $a=1$ for a fermions mass ($a=1/2$ for a
  boson mass) is recovered in the large-$N$ limit (like typically for
  the GN model~\cite{JLGN} and the scalar $\phi^4$
  model~\cite{jlprd}), this is not the case here for the NLSM, where
  $a\to 1/4$ for $N\to \infty$.}. 

With the exponent $a$ determined, one can write the resulting one-loop
RGOPT expression for the NLSM pressure as

\begin{equation}
P^{\rm RGOPT}_{ 1L} =  - \frac{(N-1)}{2} I_0^{\rm r}(m,T)+ (N-1)
\frac{m^ 2}{(4\pi) g b_0}.
\label{P1L}
\end{equation}
In the same way, the two-loop standard PT result obtained in the
previous section gets modified accordingly to yield the corresponding
RGOPT pressure at the next order of those approximation
sequences. After performing the  substitutions given by
Eq.~(\ref{delta}),  with $a=\gamma_0/b_0$ within the two-loop PT
pressure Eq.~(\ref{RenP}),   expanding now to first order in $\delta$,
next taking the limit $\delta\to 1$, gives
\begin{eqnarray} 
P^{\rm RGOPT}_{ 2L} &=& -\frac{(N-1)}{2} I_0^{\rm r}(m,T) +  (N-1)
\left(\frac{\gamma_0}{b_0}\right ) m^2 I_1^{\rm r}(m,T)\nonumber \\ &
-&  g (N-1) \frac{(N-3)}{8}  m^2 \left[I_1^{\rm r}(m,T)\right]^2 \nonumber \\ &+&
\frac{(N-1)}{4\pi} \frac{   m^2}{g \, b_0 }\left (1-
\frac{\gamma_0}{b_0}\right ) .
\label{rgopt2L}
\end{eqnarray}

\section{RG Invariant Optimization and the Mass Gap}  
\label{sec5}

To obtain the RG invariant optimized results, as a general recipe at a
given order of the ($\delta$-modified) expansion,  one expects a
priori to solve the mass optimization prescription (dubbed  MOP
below), Eq.~(\ref {pms}), and the reduced RG relation, Eq.~(\ref
{RGr}), simultaneously, thereby determining the {\it optimized} $m
\equiv {\bar m}$ and  $g \equiv {\bar g}$ ``variational" fixed point
values~\cite{JLGN,JLQCD2}.  However, at the lowest nontrivial
$\delta^0$ order,  applying  the reduced RG operator (\ref{RGr}) to
the ($\delta$-modified) one-loop pressure according to (\ref{delta})
with (\ref{exponent}),  gives a vanishing result, by
construction. Therefore, the only remaining constraint that one can
apply at this lowest order is the MOP,  Eq.~(\ref{pms}). 

 \subsection{One-loop RGOPT mass gap and pressure}

 Considering thus the MOP, Eq.~(\ref{pms}), as giving the mass as a
 function of  the other parameters $g, T, M$, it gives a gap equation
 for the optimized mass ${\bar m}(T)$, 

\begin{equation}
f_{MOP}^{(1L)}=1- 2\pi g b_0 I_1^{\rm r}(m,T) \equiv 0,
\label{gap}
\end{equation}
 or more explicitly, defining $\bar x\equiv \bar m/T$, and using
 Eq.~(\ref{I1r}),

\be 
\ln\frac{\bar m}{M}+2J_1({\bar x})+\frac{1}{b_0\, g(M)} = 0 .
\label{gap1}
\ee 
{}For $T \ne 0$ Eq.~(\ref{gap1}) gives an implicit function of
$\bar m$, due to the  nontrivial $m$-dependence in the thermal
integral.  At $T=0$, Eq.~(\ref{gap1}) immediately leads to 

\begin{equation}
{\bar m}(0) = M \exp\left( -\frac{1}{b_0\,g(M)} \right).
\label{mass1L}
\end{equation}
 It is instructive to remark that the above optimized mass
 gap is dynamically generated by the (nonlinear) interactions and
 reflects dimensional transmutation, with nonperturbative coupling dependence. 
 Accordingly the formerly perturbative Goldstone bosons
 get a nonperturbative mass, 
 indicating the restoration of the full $O(N)$ symmetry (although within our limited one-loop RG approximation, 
 at least at the same approximation level as the large-$N$ limit~\cite{ZJ,warringa}).
 The Eq.~(\ref{mass1L}) moreover fixes the optimized mass $\bar m$ to be fully
 consistent with the running coupling $g(M)$ as described by the usual
 one-loop result,

\begin{equation}
g^{-1}(M) = g^{-1}(M_0) + b_0 \ln\frac{M}{M_0} ,
\label{run1}
\end{equation}
 in terms of an arbitrary reference scale, $M_0$. 

Now replacing the mass gap expression, Eq. (\ref{gap1}), within the
one-loop pressure,  Eq. (\ref{P1L}), leads to a more explicit and
rather simple expression. Namely,

\begin{eqnarray}
P^{\rm RGOPT}_{1L} &=& -\frac{(N-1)}{\pi} T^2 \left\{
J_0\left(\frac{\bar m}{T}\right) \right. \nonumber\\ &+&
\left. \frac{1}{8} \left(\frac{\bar m}{T}\right)^2 \left[ 1+4
  J_1\left(\frac{\bar m}{T}\right)\,\right]\right\}  ,
\label{P1Lfin}
\end{eqnarray}
where  $\bar m \equiv \bar m(g,M,T)$  is given by the solution of
Eq.~(\ref{gap1})~\footnote{  Notice that the term proportional to
  $g^{-1}$ in Eq.~(\ref{P1L}) has been absorbed upon using
  Eq.~(\ref{gap1}), such that there  are no particular problems for
  $g\to 0$ in Eq.~(\ref{P1Lfin}).}.

{}For completeness, we also give the corresponding pressure at
zero-temperature, that we  will use to subtract from
Eq.~(\ref{P1Lfin}) in the numerical illustrations to be given in
Sec.~\ref{sec6}, such as to obtain a conventionally normalized
pressure, $P(T=0)\equiv 0$.  {}From Eq.~(\ref{P1Lfin}) one
obtains

\be 
P^{\rm RGOPT}_{1L}(T=0) = -\frac{(N-1)}{8\pi} \,\bar m^2(0) ,
\label{P1LT0}
\ee 
where the $T=0$ mass-gap is given in Eq.~(\ref{mass1L}).

{}From the above expressions, one may anticipate that
  Eqs.~(\ref{gap1}) and (\ref{P1Lfin})  exhibit ``exact" scale
  invariance (of course exact upon neglecting higher order terms at
  this stage), as it will be further illustrated by the numerical
  evaluations performed in Sec.~\ref{sec6}.

\subsection {Large-$N$ mass gap and pressure}
\label{secLN}

Before deriving the RGOPT results at the next (two-loop) order, for
completeness we consider  the large-$N$ (LN) limit of the model, as it
can be directly obtained from the previous one-loop RGOPT result and
it will be also studied for comparison purposes in the sequel.  The LN
limit is straightforwardly generated from the usual procedure of
rescaling the coupling as 

\be 
g \equiv g_{LN}/N,
\label{LNresc}
\ee 
and then taking the limit $N\to\infty$ in the relevant expressions
above, such that typically
any $(N-1)$, $(N-2)$, $\cdots$  factors in Eq.~(\ref {P1L}), or
previous related expressions reduce to $N$, while higher orders
terms are $1/N$-suppressed~\footnote{One must note one subtlety:  the
  second term of Eq.~(\ref{RenP}), formally not vanishing after using
  (\ref{LNresc}) for $N\to\infty$, should not be included,  as it
  would be double-counted, since such term (and all higher order LN
  terms) is actually consistently generated from using the LN limit of
  the mass gap Eq.~(\ref{gap1}) within the LN pressure
  Eq.~(\ref{PLN}).}. Therefore, the  LN limit of the RGOPT pressure
expression (\ref{P1L}) takes the explicit form 

\be 
P^{\rm LN} =  - \frac{N}{2} I_0^{\rm r}(m,T)+ N \frac{m^ 2}{2
  g_{LN}}.
\label{PLN}
\ee 
Note that Eq.~(\ref{PLN}) is fully consistent with the first order
of the nonperturbative  two-particle irreducible (2PI) CJT
formalism~\cite{CJT} result given in Ref.~\cite{giacosa} (the
Eq.~(2.42) in that reference), upon further subtracting  $P(T=0)$ from
Eq.~(\ref{PLN}),  with the large-$N$ limit of the mass gap $\bar m$,
Eq.~(\ref{mass1L}),  $b_0\, g \to g_{LN}/(2\pi)$,  also consistent
with the Eq.~(2.44) of Ref.~\cite{giacosa}. The authors of
Ref.~\cite{giacosa} have explained the reasons for the strict
equivalence of their first order CJT approximation with the large-$N$
results~\cite{warringa}. 

  The similarities
between RGOPT at one-loop order and the first nontrivial order of 2PI
results were already noticed in the context of the scalar $\phi^4$
model~\cite{jlprd}.   Hence, at large-$N$ the rather simple
  RGOPT lowest (one-loop) order procedure is equivalent to resumming
  the leading order temperature dependent terms for the mass
  self-consistently and this result remains valid at  any
  temperature. 
  
Note also that within the standard nonperturbative LN calculation framework, the last term in Eq.~(\ref{PLN}) 
arises when this approximation is implemented with, e.g., 
the traditional auxiliary field method \cite {warringa},  and is crucial  
to maintain consistent RG properties. 
Accordingly, the LN pressure will also exhibit ''exact`` scale invariance, at this approximation level. 
However, as we will examine explicitly
below, the pressure at the NLO in the $1/N$ expansion, although an a priori more precise nonperturbative quantity, 
is not exactly scale invariant, exhibiting a moderate residual scale dependence. 

The LN pressure (\ref{PLN}) now scales as $\sim
N$. Thus,  to compare this LN approximation  in a sensible manner with
the true physical pressure (i.e. for a given physical $N_{phys}$
value),  in the numerical illustrations to be performed in
Sec.~\ref{sec6}, we will adopt the standard convention to take the
overall $N$ factor of the  LN pressure (\ref{PLN}) as $N\to N_{phys}$
(typically $N_{phys}=3,4,\cdots$ in our numerical illustrations).

\subsection{Two-loop RGOPT mass gap and pressure}

Going now to two-loop order, the mass optimization criterion
Eq.~(\ref{pms}) applied to the RGOPT-modified two-loop pressure
Eq.~(\ref{rgopt2L}) can be cast, after straightforward
algebra, in the form (omitting some irrelevant overall factors):
\bea 
& f_{MOP}^{(2L)}\equiv \left\{ 3N-5 -b_0 (N-3)\, g \,\left[1 +Y
  +2x^2 J_2(x)\right]\right\}   \nn \\ &\times m \,\left(\frac{1}{b_0\,g}
+Y \right)   = 0,
\label{pms2loop}
\eea   
where, we have defined for convenience the following dimensionless quantity
(compare with Eq.~(\ref{I1r})),
\be 
Y \equiv \ln \frac{m}{M}+2J_1(x) = -2\pi\,I_1^r(m,T),
\label{Ydef}
\ee 
and the thermal integral, $x J_2(x)\equiv \partial_x J_1(x)$ reads
\begin{equation}
J_2(x) = \int_0^\infty dz
\frac{[e^{\omega_z}(1+\omega_z)-1]}{\omega_z^3 (1-e^{\omega_z})^2} .
\end{equation}
Alternatively, the reduced RG equation (\ref{RGr}), using the exact
two-loop $\beta$-function Eq.~(\ref{beta}), yields
\bea 
& f_{RG}^{(2L)} \equiv  m^2\,\left[ g \frac{(3N-5)}{2\pi} +(N-3) \times \right. \nn \\ 
& \left. \left\{ 1+\frac{N-2}{\pi} g \,Y \left[1+ \frac{N-2}{4\pi} g\,
  (1+\frac{g}{2\pi})\,Y\right]\right\} \right]=0 .
\label{rg2loop}
\eea 

When considered as two alternative (separate) equations, 
(\ref{pms2loop}) and (\ref{rg2loop}), apart from having the trivial solution $\bar m=0$, 
also have a more interesting nonzero mass gap solution, $\bar
m(g,T,M)$, with nonperturbative dependence on the
coupling $g$.  It is convenient to solve Eq.~(\ref{rg2loop}) first
formally as second-order algebraic equations for $Y$, as function of
the other parameters, and solving (numerically) the mass gap
$\bar m(g,T,M)$ using Eq.~(\ref{Ydef}).  To get more insight on those
implicit self-consistent equations for $m(g,T)$, let us first observe that the MOP
Eq.~(\ref{pms2loop}) factorizes, with the first factor  recognized as
the one-loop MOP Eq.~(\ref{gap1}). Now it is easily seen that the
other nontrivial solution, given by cancelling the second factor in
Eq.~(\ref{pms2loop}), gives at $T=0$ a behavior of the coupling (or
equivalently, of the mass gap), asymptotically of the form, when $M\gg
m$,
\be 
g(m,M,T=0)\stackrel{M\gg m}{\longrightarrow}   \left
(\frac{5-3N}{N-3} \right )\,\frac{1}{b_0\ln \frac{M}{m}} ,
\label{nonAF}
\ee 
which badly contradicts the asymptotic freedom (AF) property
of the NLSM,  the coefficient of the right-hand-side of
Eq.~(\ref{nonAF}) having the opposite sign of AF for any $N>3$. We
therefore unambiguously reject this  solution~\footnote{This
  illustrates another advantage of the RGOPT construction, namely that
  by simply requiring~\cite{JLQCD2} AF-compatible branch solutions for
  $g\to 0$ one often can select  a unique optimized solution at a
  given perturbative order.}, which means that at two-loop order, the
correct $g\to 0$ AF-compatible physical branch solution of
Eq.~(\ref{pms2loop}) for the mass gap is unique and formally  the same
as the one-loop solution from Eq.~(\ref{gap1}).  But more generally one expects
both the mass optimization and the RG solution of (\ref{rg2loop}) 
to differ quite drastically from the one-loop solution, obviously since incorporating higher order RG-dependence.
For the NLSM one also immediately
notices that the case $N=3$ is very special, as could be expected, since the
two-loop original perturbative contribution in Eq.~(\ref{RenP})
vanishes. Once performing the $\delta$-expansion at two-loop order, even if that gives extra terms, 
as can be seen by comparing Eqs.~(\ref{P1L}) and (\ref{rgopt2L}), these also vanish for $N\to 3$, 
since $\gamma_0(N=3)=0$, see Eq.~(\ref{gamma0}).
Moreover, if using only Eq.~(\ref{pms2loop}), it reduces for $N=3$ to
the last factor, identical to the one-loop MOP Eq.~(\ref{gap1}). Thus, if using the latter MOP prescription for
$N=3$, one would only recover the one-loop mass gap solution
Eq.~(\ref{gap1}), which implies no possible improvement from one- to two-loop order.  
However, using instead the {\em full} RG equation Eq.~(\ref{RG}), as we will specify below, 
gives a nontrivial two-loop mass gap solution 
$\bar m(g)$ that is intrinsically different and goes beyond the one-loop solution Eq.~(\ref{gap1}) even for $N=3$.
It thus implies that the final RGOPT pressure, considered as a function of the coupling, $P(\bar m(g))$, 
will nevertheless be different from the one-loop Eq.~(\ref{P1Lfin}). In that way, even for $N=3$,
where the purely perturbative two-loop term cancels, the RGOPT
procedure allows a different (and a priori improved) approximation from
one- to two-loop order. We will see that those differences between one- and two-loop RGOPT expressions 
happen to be maximal
for $N=3$ (which is intuitively expected since it is the lowest possible physical value for the interacting NLSM). 
This is a quite sensible feature in view of the fact
that we will compare the RGOPT one- and two-loop results with
nonperturbative lattice simulations~\cite{giacosa}, which  are,
however, only available for $N=3$ at present. 

Now for $N>3$,  in principle it would be desirable to  find a
simultaneous (combined) solution of Eqs.~(\ref{pms2loop}) and
(\ref{rg2loop}), such as to obtain the approximate optimal  ``fixed
point"   set $\{{\bar m},{\bar g}\}$,  as was done in some $T=0$ models~\cite{JLGN,JLQCD1,JLQCD2,JLqq}.   
{}For $T\ne 0$ it
would leave a given pair $(T,M)$ as the only input parameters.  But
for the present NLSM, a rather unexpected feature happens: as easily
derived, e.g., by solving the correct AF physical solution of
Eq.~(\ref{pms2loop}) first for $Y$, and substituting in
Eq.~(\ref{rg2loop})), after some straightforward algebra the latter
readily reduces to
\be 
g\,m^2 =0 .
\label{rg+opt}
\ee 
Thus, at two-loop order there is {\em no} such
nontrivial RG and MOP combined solution in the NLSM.  This is
not an expected result in general for other models, but that one can easily trace back to the
specific renormalization properties of the NLSM vacuum energy in $\ms$-scheme as discussed in Sec. \ref{sec3}.
(Equivalently the subtraction coefficient $s_1$ in Eq.~(\ref{s1}) vanishes due to the
the peculiar NLSM $b_1=2\gamma_1$ relations between two-loop RG coefficients for any
$N$~\footnote{More precisely, if $s_1\ne 0$, as it happens in other models, the
  two-loop MOP Eq.~(\ref{pms}) does not factorize like in
  Eq.~(\ref{pms2loop}) with  the one-loop solution factor. Therefore,
  a nontrivial combined solution does exist.}. 
  It is thus a peculiar feature of the
  NLSM, unlikely to occur in a large class of other
  models. It simply means that at two-loop order the NLSM pressure has a too simple 
  $(m, g)$  dependence to provide such a nontrivial intersecting optimal solution
  of the two relevant, RG and MOP equations.
Nontrivial
combined RG and MOP solutions should most likely exist for the NLSM at the next three-loop
order, which is  however beyond the scope of the present analysis.
Therefore, restricting ourselves to the two-loop order for simplicity,
for $N>3$ we have to select {\em either} Eq.~(\ref{pms2loop}), {\em or} Eq.~(\ref{rg2loop})
to give the mass gap, then fixing the coupling more conventionally
from its more standard perturbative behavior. 
Besides these peculiar NLSM features, the latter prescription is also  more transparent 
to compare with former similar SPT or HTLpt available results for
other models, where  the (mass) optimization or other used prescriptions only provide a
mass gap as a function of the coupling, and the coupling is not fixed
by other procedures, thus generally chosen as dictated by the standard
(massless) perturbative behavior~\cite{SPT,htlpt1,HTLPT3loop}. 

Now Eq.~(\ref{rg2loop}) alone happens to have real solutions only at
large-$N$, in contrast with Eq.~(\ref{pms2loop}), which has real
solutions for any $N >2$. But since the correct NLSM AF branch of the mass
optimization solution of Eq.~(\ref{pms2loop}) behaves accidentally very much
like at one-loop order, as explained above, we do not expect to  gain much from it when going to
the two-loop order.  A more promising alternative is to use instead
the {\em complete} RG equation, which combine Eq.~(\ref{pms2loop}) and Eq.~(\ref{rg2loop})
in the form (omitting irrelevant overall factors):
\begin{equation}
f_{full \,RG}^{(2L)} \equiv f_{RG}^{(2L)} + 2m\,\gamma_m f_{MOP}^{(2L)} =
0, 
\label{RGfull}
\end{equation} 
where $\gamma_m$ consistently includes two-loop  ${\cal O}(g^2)$
terms, see Eq.~(\ref{gammam}).  Not only this contains the maximal 
RG ``information", in contrast with the simpler mass optimization
Eq.~(\ref{pms2loop}), but since the RG equation is considered alone,
ignoring its possible combination with the mass optimization
(\ref{pms}), Eq.~(\ref{RGfull}) is more appropriate than the
reduced RG Eq.~(\ref{rg2loop}), which is obtained only after using
Eq.~(\ref{pms}).  The input parameters are now  $\{T,M,g(M)\}$  and
Eq.~(\ref{RGfull}) allows to fix  $\bar m$.  {}Moreover,
Eq.~(\ref{RGfull}) does give real solutions for any value of $N$ and
for small to moderately large  couplings.  In addition, as mentioned above, for the very special
case $N=3$ it also gives a mass gap solution that is intrinsically different from the one-loop
solution (\ref{gap1}), a very welcome feature. (This happens because of the additional 
coupling dependence within $\gamma_m$ in Eq.~(\ref{RGfull}), which turns the otherwise
trivial RG solution of Eq.~(\ref{rg2loop}) alone, for $N=3$, into a nontrivial solution). 
All the previous considerations therefore impose
Eq.~(\ref{RGfull}) as the most sensible and unique prescription, that we follow from now on.
{}For very strong couplings
and $N>3$ 
the solutions of (\ref{RGfull}) become complex, nevertheless, if needed, this region can
still be explored by solving the less stringent condition given by
Eq.~(\ref{pms2loop})~\footnote{When none of the RG and MOP equations
  give real solutions, one may attempt an additional
  prescription, performing perturbative renormalization scheme
  changes, which may recover real solutions,
  as was done at $T=0$ in Ref.~\cite{JLQCD2}.  The
  generalization of this extra procedure to $T
  \ne 0$ in the present context appears however numerically more involved
  and beyond the scope of the present paper.}. 

We emphasize that when using Eq.~(\ref{RGfull}) to determine the optimized
pressure results, one may consider that the coupling runs in the way dictated
by the standard perturbative approximation. Hence, apart from the one-loop running in Eq.~(\ref{run1}), we
also need the two-loop running coupling, with exact expression given e.g. in ~\cite{jlprd}, which can be
approximated as follows with sufficient accuracy (as long as $g$
remains rather moderate $g\sim {\cal O}(1)$),
\begin{eqnarray} 
g^{-1}(M) &\simeq & g^{-1}(M_0) +b_0 L   +(b_1 L)g(M_0) \nonumber
\\  &-&\left (\frac{1}{2} b_0 b_1 L^2\right )  g^2(M_0) \nonumber
\\  & -&\left (\frac{1}{2} b_1^2 L^2 -\frac{1}{3} b_0^2 b_1 L^3\right
) g^3(M_0)  \nonumber \\ &+& {\cal O}(g^4), 
\label{run2}
\end{eqnarray} 
where $L = \ln (M/M_0)$.  As it is standard,  one can compare the
  scale dependence of the different approximations by setting
  $M=\alpha M_0$, where $M_0$ is an arbitrary reference scale, and
  varying $\alpha$ in a given range from  Eq.~(\ref{run1}) or
  Eq.~(\ref{run2}), at one- and two-loop orders, respectively. 

\subsection{Comparison with PT pressure and Debye screening pole mass}
\label{secPT}

In  order to compare our results with the standard (massless) perturbation theory
(PT) in the present NLSM model, it is appropriate to  consider the
high-$T$ expansion of the relevant expressions. This is also relevant
for a (merely qualitative) comparison with HTLpt results~\cite{htlpt1}
in other models, since the latter proceeds with expansions in powers
of $x=m/T$. We will see in this subsection that  Eqs.~(\ref{gap1}), or
equivalently, Eqs.~(\ref{pms2loop}) and (\ref{RGfull}) at
two-loop order, have relatively  simple mass gap solutions given in
this case as a systematic perturbative expansion in powers of the
coupling.

It is thus useful to consider  the well known high-$T$ expansions,
where $x=m/T\ll 1$,  for the thermal  integrals~\cite {kapusta},
\begin{eqnarray}
&&J_0(x)= - \frac{\pi^2}{6} + \frac{\pi}{2} x+ \left ( \frac{x}{2}
  \right)^2  \left [ \ln \left(\frac{x e^{\gamma_E}}{4\pi} \right )
    -\frac{1}{2} \right ] +{\cal O}(x^4), \nonumber \\
\label{J0exp}
\\ &&J_1(x)= -\frac{\pi}{2x} -\frac{1}{2}\, \ln \left(\frac{x
  e^{\gamma_E}}{4\pi} \right ) +{\cal O}(x^4).
\label{J1exp}
\end{eqnarray}
We also introduce the Stefan-Boltzmann (SB) limit of the renormalized
NLSM pressure, which will enter as a reference pressure in many of the
numerical examples to be given below  in Sec.~\ref{sec6},
\begin{equation}
P_{\rm SB} = (N-1) \frac{\pi}{6} T^2 ,
\label{SB}
\end{equation} 
which is obtained by taking the massless, or high-$T$ limit, of
Eq.~(\ref{J0exp}) of the one-loop perturbative result Eq.~(\ref{P0}).
This gives for the one-loop RGOPT pressure Eq.~(\ref{P1Lfin}),
\begin{equation}
\frac{ P^{\rm RGOPT}_{1L}}{P_{SB}} = 1 - \frac{3}{\pi} {\bar x} -
\frac{3}{2\pi^2} {\bar x}^2  \left [L_T - \frac {1}{b_0\, g}  \right ]
+ {\cal O}({\bar x}^4) ,
\label{optHT}
\end{equation}
where we have defined $L_T= \ln [ M e^{\gamma_E}/(4\pi T)  ]$. 

Next, the optimized mass solution, obtained from Eq.~(\ref{gap1}),
can be expressed as function of the coupling:
\begin{equation}
\bar x \equiv \frac{\bar m}{T} = \frac{\pi\,b_0 g(M)}{1-b_0 g(M)\,L_T} ,
\label{mbarHT}
\end{equation} 
or that simply gives, when expanding to the lowest perturbative order, 
\begin{equation}
\bar m (T) \simeq  \pi\,b_0\,g(M) T +{\cal O}(g^2).
\label{mbarHT2}
\end{equation}
Note that using the optimized mass gap solution (\ref{mbarHT}) within
Eq.~(\ref{optHT}), the latter takes a much simpler expression (in the
high-$T$ limit here considered),
\begin{eqnarray}
& & \frac{ P^{\rm RGOPT}_{1L}}{P_{SB}} = 1 - \frac{3}{2\pi} {\bar x}  +
     {\cal O}({\bar x}^4)   \nonumber \\ & & \simeq 1-\frac{3}{2}\,
     \frac{b_0 g(M)}{1-b_0 g(M)\,L_T}  = 1-\frac{3}{2}\,
     b_0 g(\frac{4\pi T}{e^{\gamma_E}}) , 
\label{optHTsimp}
\end{eqnarray}
using Eq.~(\ref{run1}) in the last term.
  At the next two-loop order, one obtains 
  in the high-$T$ approximation a relatively compact expression of the RGOPT pressure Eq.~(\ref{rgopt2L}):
\bea
&& \frac{ P^{\rm RGOPT}_{2L}}{P_{SB}} = 1-\frac{3\bar x}{4\pi}(\frac{3N-5}{N-2})(1-\frac{\bar x}{(N-2)\,g}) \nn \\
&&  -\frac{3\bar x^2}{4\pi^2}(\frac{N-1}{N-2})\,L_T -\frac{3g(N-3)}{16\pi^3}\left(\pi +L_T\,\bar x\right)^2 ,
\label{rgopt2Lht}
\eea
  where the correct (i.e. AF) solution $\bar m\equiv \bar x T$ of the full RG
  Eq.~(\ref{RGfull}) is given simply by one of the roots of a quadratic equation.
  $\bar m$ is in general different from the one-loop solution 
  (\ref{mbarHT}), as expected since it now involves two-loop order RG coefficients
  $b_1, \gamma_1$: indeed it has a rather involved
  dependence on $g$ that we refrain to give explicitly. But once perturbatively re-expanded, 
  it coincides at first order with (\ref{mbarHT}) (for any $N>3$), which is a nontrivial 
  perturbative consistency check of our construction.
   Replacing this exact two-loop $\bar m$ as a function of $g$ within the two-loop
  pressure (\ref{rgopt2Lht}), one obtains
  an expression that differs from the one-loop pressure, Eq.~(\ref{optHTsimp}), by higher order perturbative terms, starting
  at ${\cal O}(g^3)$:
\bea 
\frac{ P^{\rm RGOPT}_{2L}}{P_{SB}} &=& 1-\frac{3}{2}\, \frac{b_0 g(M)}{1-b_0 g(M)\,L_T} 
 \nn \\ &  + & \frac{3\,g^3(M)}{\pi^3}\,\frac{(N-2)^4}{(N-3)^2(3N-5)}  +{\cal
  O}(g^4) ,
\label{rg2HT}
\eea  
valid strictly only for $N> 3$.
Accordingly the extra terms in Eq.~(\ref{rg2HT}) illustrate rather simply (perturbatively) 
the additionnal contributions from two-loop RGOPT with respect to the one-loop pressure (\ref{optHTsimp}). 
One should keep in mind, however, that the
exact two-loop pressure (\ref{rgopt2L}), including full $g$- and $T$-dependence from the exact $\bar m(g,T)$ 
(which we illustrate numerically below mainly for $N=4$) 
has a much more involved, nonperturbative dependence on the coupling (and temperature).
 In particular, while Eq.~(\ref{rg2HT}) is a relatively good approximation for  moderate $g$ and $N\ge 4$,
it is not valid for the very special case $N=3$, for which the exact pressure $P(\bar m(g))$ actually does not show 
any singular behavior, since both the original expression Eq.~(\ref{rgopt2Lht}) and $\bar m(g)$ are regular for $N=3$. 
The singularity at $N=3$ seen in Eq.~(\ref{rg2HT}) is thus unphysical, being only an artifact of having  
perturbatively reexpanded $\bar m(g)$,  
which exhibits a $1/(N-3)$ terms at order $g^2$. 
(Therefore for the case $N=3$ some care is needed in the numerics to take the limit $N\to 3$ {\em before} possibly expanding in
perturbation, which is however not needed). 
Still, Eq.~(\ref{rg2HT}) indicates crudely that the difference between one- and two-loop RGOPT should be maximal for
$N=3$, which is also true for the exact (regular) expression, and was intuitively expected, 
since $N=3$ is the lowest physically nontrivial 
value. At the other extreme for $N\to \infty$, one can easily check that 
the two-loop optimized mass and pressure tend  
towards the corresponding one-loop quantities, i.e. the LN results like the pressure Eq.~(\ref{PLN}).  \\

To compare the previous RGOPT results with the
standard PT ones,  one can start by deriving the PT pressure directly
from Eq.~(\ref{RenP})  in the massless limit, which at this two-loop
order is well defined. It gives
\begin{equation}
\frac{ P^{\rm PT}}{P_{SB}} = 1 - \frac{3}{2} \, \gamma_0\, g(M) +{\cal
  O}(g^2). 
\label{Ppt}
\end{equation}

Another quantity of interest is the purely perturbative thermal
Debye (pole) mass:  at one-loop order it can be derived starting from
the self-energy~\cite{nlsmrenorm,ZJ},
\begin{equation}
\Gamma^{(2)}(p^2) = p^2 (1+g_0 I_1) +m^2_0 \left [1+\frac{(N-1)}{2}
  g_0 I_1\right ]+{\cal O}(g^2)\,,
\label{Gam2}
\end{equation}
where $I_1$ is the (Euclidean) one-loop integral given by
Eq.~(\ref{I1def}) in $\overline {\rm MS}$ renormalization scheme, with
the thermal part $J_1(x)$ having the high-$T$ expansion (\ref{J1exp}).
Taking, thus, the pole mass $p^2\equiv -m^2_D$ in Eq.~(\ref{Gam2}),
after mass renormalization, it gives, in the massless limit $m\to 0$
relevant for the pure thermal one-loop  mass, the result
\begin{equation}
{m^{1L}_D} = \frac{N-3}{8}\: g(M)\:T \equiv \pi \, \gamma_0 
\,g(M)\:T .
\label{mDB}
\end{equation}
Equation~(\ref{mDB}) can be contrasted with the more nonperturbative
RGOPT result (\ref{mbarHT2}).  Accordingly, note that the RGOPT
optimized mass (\ref{mbarHT2}) appears to have  a different
perturbative behavior than the  Debye pole mass (\ref{mDB}):
Namely, with $\gamma_0\to b_0$, and similarly for the pressure,
comparing the RGOPT result Eq.~(\ref{optHT})  with the PT pressure
(\ref{Ppt}), the slopes of both masses at the origin as a function of
$g$ are different.  However, one should not be surprised by these
differences. {}First, in contrast with the physical one-loop Debye
(pole) mass,  the optimized mass (\ref{mbarHT2}) is only an
intermediate unphysical quantity in the optimization  procedure, aimed
to enter the final pressure to make it a (nonperturbative) function of
$g$ only.  Second, the resulting $P^{\rm RGOPT}(g)$, obtained by such
a construction, has a priori more nonperturbative content. Thus, it
has  no reason to generate a function that exactly matches the one
generated by the standard PT.  This is similar 
  to the fact that the pressure, in the nonperturbative
  LN approximation, also  is
  a function of $g$ that is intrinsically different from the purely PT
  pressure,
\be 
\frac{ P^{\rm LN}}{P_{SB}} = \frac{N}{N-1}\left(1 - \frac{3}{2} g_{LN} \right)
+{\cal O}(g^2_{LN}) .  
\ee 
Thus, at this stage the coupling value is
an essentially arbitrary input, being not fixed from a physical input
at a given scale, in both PT, LN and RGOPT approximation cases, and a
physical input would fix a priori different values of the coupling in
different approximation schemes. However, the perturbative
and physical consistency of the RGOPT pressure result can be checked
by appreciating that, once the {\em arbitrary} mass in
Eq.~(\ref{optHT}) is replaced with the {\em physical} thermal mass
$m_D$ Eq.~(\ref{mDB}), one consistently recovers the  standard PT
pressure as function of $g$, Eq.~(\ref{Ppt}).  In other words, when
expressed in terms of physical quantities (like here the Debye
pole mass) the RGOPT results  are consistent with standard PT for
$g\to 0$  (see, e.g., Ref.~\cite{jlprd} for a detailed discussion of
similar results for the scalar $\phi^4$ model). 

\subsection{Comparison with standard SPT/OPT}

{}For completeness, let us review how the more standard OPT (or SPT)
approximation is obtained and derive it for the NLSM, for useful 
comparison purpose with the  RGOPT results.  In this case, one
starts back again with Eq.~(\ref{press2l}),  but as already emphasized
above, there are two important differences with the previously derived
RGOPT construction. {}First, the standard
SPT/OPT, as was considered in various models, generally ignored the finite
vacuum energy subtraction terms like in Eq.~(\ref{skdef}),
required to restore the perturbative RG invariance as we have
discussed. The second difference  regards the Gaussian term when
performing  the interpolation, Eq.~(\ref{delta}), since in the
standard OPT case the exponent $a$ is fixed  in an {\it ad hoc} way as
$a=1/2$. While it should be clear from the above RGOPT construction  that such
prescription will therefore lack explicitly RG invariance, we nevertheless follow
exactly the procedure as it was applied in various other models, to illustrate the
differences in properties of corresponding thermodynamical quantities
as compared to the RGOPT, in particular concerning their residual
scale dependence.

It is thus straightforward to obtain the SPT/OPT two-loop pressure
from the RGOPT result, Eq.~(\ref{rgopt2L}): upon first omitting the
finite contribution to ${\cal E}_0$, given by the last term in
Eq.~(\ref{rgopt2L}), furthermore upon replacing the RGOPT exponent $a=\gamma_0/b_0$
by the standard $a=1/2$  in the second term. These modifications lead
to 
\begin{eqnarray}
P^{\rm SPT}_{2L}&=&  -\frac{(N-1)}{2} I_0^{\rm r}(m,T) +
\frac{(N-1)}{2}  m^2 I_1^{\rm r}(m,T)  \nonumber \\ &-& g (N-1)
\frac{(N-3)}{8}  m^2 \left[I_1^{\rm r}(m,T)\right]^2 .
\label{Pspt}
\end{eqnarray}
One should first appreciate that, contrary to the RGOPT case, the
SPT/OPT does not provide a non-trivial (i.e. coupling-dependent) 
optimized mass gap result when only the (one-loop)  quasi-particle
contribution, given by the first two terms on the right-hand-side of
Eq.~(\ref{Pspt}), is accounted for. This is a general feature,
not specific to the NLSM. Applying thus the
mass optimization  Eq.~(\ref{pms}) to the complete two-loop $P^{\rm
  SPT}_{2L}$ result, one  obtains  a second-order equation quite
analogous to Eq.~(\ref{pms2loop}),
\begin{equation}  
f_{SPT}^{(2L)} = \left[ 1 - \gamma_0 \, g (1 + Y) \right] - 2 \gamma_0
\, g x^2 J_2(x) = 0 ,
\end{equation} 
where the quantity $Y$ was defined in Eq.~(\ref{Ydef}).  However, this
equation has no real solution for any $N$.  Moreover,  upon taking
  its high-$T$ approximation, it gives an unphysical 
  solution, as it no longer depends on the mass. This last feature is an
unusual situation within the SPT/OPT/HTLpt applications since, at
least for other models considered in the literature, these
approximations often provide real results at the first nontrivial
order. And, in particular, they usually recover the LN result when $N
\to \infty$~\cite{optLN} as,  for example, in the case of the $\lambda
\phi^ 4$ scalar theory~\cite{optON,OPT3l}.  In this
situation, a frequently used alternative prescription to nevertheless
define a mass in SPT (or similarly HTLpt)~\cite{HTLPT3loop} is
to employ the purely perturbative NLSM Debye pole  mass, as given by
Eq.~(\ref{mDB}).  Since we are interested in the complete temperature
range and not solely in the high-$T$ regime,  we could rather
  derive    ${\bar m}_{SPT}$ as the solution of the {\em full}
  one-loop self-consistent mass gap equation obtained from
  Eq.~(\ref{Gam2}), 
\begin{equation}
\bar m \equiv \lim_{m\to 0} m\left[1 + 2\pi \gamma_0\,g\, I_1^{\rm
    r}(\bar m,T) +{\cal O}(g^2) \right] .
\label{gapspt}
\end{equation}
But, at this perturbative order,  the physical solution of
Eq.~(\ref{gapspt}) is nothing but the high-$T$ one-loop  Debye 
mass $m_{D}$, already given in Eq.~(\ref{mDB}). 

{}For completeness and later use, we also give the expression of the
two-loop SPT pressure  Eq.~(\ref{Pspt}) in the high-$T$ approximation,
upon using its mass-gap solution Eq.~(\ref{mDB}), therefore becoming
only a function of  $g(M)\sim g(T)$,

\be 
\frac{P^{\rm SPT,high T}_{2L}}{P_{SB}} = 1-\frac{3}{2} \gamma_0 g
-\frac{3}{2} \gamma_0 g  \left(1+\gamma_0 g L_T\right)^2 .
\label{Psptht}
\ee 
In particular, the previous expression is more appropriate for a
(very qualitative) comparison with the two-loop HTLpt QCD (pure
gluodynamics) pressure \cite{HTLPT3loop},  which is 
only available in the small  $m/T$ (high-$T$) expansion approximation.

\section{Numerical Results}
\label{sec6}

Before proceeding to numerical comparisons of the different
approximation methods previously considered,  we should specify how to
fix the relevant input parameters, which we discuss next.
%
\subsection{Input parameter choice}

As already mentioned, at this stage the coupling $g$ in all previous
RGOPT, PT, SPT approximations of the pressure is to be considered an
arbitrary input.  Ideally, if we had experimental data for some
physical observable, like for other models, we  could fix $g$
typically at some scale and/or temperature. Accordingly it is clear
that the resulting $g(T)$ values  would be a priori different within
different approximations.  Specially, since the LN approximation
necessarily implies to rescaling the coupling, Eq.~(\ref{LNresc}),
the rescaled coupling $g_{LN}$ value could be substantially different
from those for other approximations, for the same observable input
given  at some physical scale for a finite $N$ value. Now,
apart from comparing with other available nonperturbative results 
(like the NLO $1/N$ expansion~\cite{warringa} or lattice simulations for
$N=3$\cite{giacosa}), our purpose is also mainly to illustrate the RGOPT scale dependence
improvement as compared to the standard PT 
and the SPT approximation.  For the latter comparison it is more sensible to compare scale
dependences of the different results for the {\em
  same} ``reference"   coupling values. But since we also compare the different
thermodynamical quantities with the LN ones, one aims  to choose
$g_{LN}$ input values in a range that is a priori comparable with
other approximations. It is clear from Eq.~(\ref{P1Lfin}) and
(\ref{PLN}) that the one-loop RGOPT and LN approximations are
essentially equivalent, only up to a rescaled coupling (\ref{LNresc}):
indeed, the correct LN result was derived from Eq.~(\ref{P1Lfin}).
Thus we find it sensible to compare the results  for a given finite
$N_{phys}$ input by taking 
\be 
g_{LN}(M_0) \simeq (N_{phys} -2) g(M_0)
\label{gLNrelation}
\ee 
When satisfying exactly this relation, the LN and (one-loop) RGOPT
describe essentially the same physics: if one would fix $g(M_0)$ for the
different approximations by comparing those to real data, one would expect
to obtain  something close to Eq.~(\ref{gLNrelation}), except for the
other difference being the $N-1\to
N$ overall factor in the LN  pressure~\footnote{Accordingly $P_{\rm
    SB}$ has a higher value in LN.  Hence, one can anticipate that the
  LN results will overestimate the {\it true} SB limit, as given by
  Eq.~(\ref {SB}), at high temperatures~\cite{giacosa}.}.  Concretely,
the numerical illustrations below will be mainly for the case of
$N_{phys}=4$, $g(M_0)=1=g_{LN}(M_0)/2$,  where $M_0$ is the arbitrary
reference scale, or for $N_{phys}=3$, $g(M_0)=g_{LN}(M_0)$ when
comparing with the lattice results. 

To investigate and compare  the scale variation behavior of the
different approximations  in our analysis below, as it is customary,
we set the arbitrary $\ms$ scale  as $M=\alpha M_0=2\pi T\alpha $ and
consider $0.5 \leq \alpha \leq 2$ as representative values of scale
variations. Note however that this formal identification of
  the arbitrary  renormalization scale $M$ with a temperature is only 
  justified strictly at high temperature~\cite{Trev}, while the genuine
  nonperturbative arbitrary $T$-dependence of the coupling is in
  general not known. Accordingly for the SPT Eqs.~(\ref{Pspt}),
  (\ref{Psptht}) and PT Eq.~(\ref{Ppt}), we impose the standard prescription 
  $g(M\sim 2\pi T)$ with the running dictated e.g. at one-loop by Eq.~(\ref{run1}). This  
  guarantees the correct SB limit of (\ref{Ppt}) and Eq.~(\ref{Psptht}) at very high $T$, and 
  is often adopted in the literature even for relatively low $T$ values. In contrast for the RGOPT pressures 
  Eqs.~(\ref{P1L}) or  Eq.~(\ref{rgopt2L}), the running $g(M\sim T)$
  as dictated by 
  Eqs.~(\ref{run1}), (\ref{run2}) is consistently embedded (although only approximately at two-loops),
  as is explicit e.g. from Eq.~(\ref{optHTsimp}) in the high-$T$ approximation. This is  
  a consequence of the (perturbative) RG invariance-restoring subtraction terms in the pressure expressions,
  and it automatically gives the SB limit at (very) high $T$.
  Alternatively, as already emphasized previously, if a nontrivial
  combined RG and MOP solution of Eq.~(\ref{pms2loop}) and
  (\ref{rg2loop}) would be available at the two-loop order in the present
  model (which unfortunately is not the case), this solution
  would effectively provide an approximate ``nonperturbative'' ansatz for the 
  $T$-dependent coupling, likely departing much from Eqs.~(\ref{run1}), (\ref{run2}) at low $T$. \\
  Having previously derived that the one-loop RGOPT (and LN
similarly) pressure is exactly scale invariant for any coupling value,
we will illustrate the moderate residual scale dependence at the
two-loop RGOPT order and those of the other approximation schemes, for
a moderately nonperturbative coupling choice, $g(M_0)\simeq 1$.  
  It is clear that due to the perturbative
  running, for a very large input coupling the scale dependence drastically
  increases (except for the one-loop RGOPT result being exactly scale
  invariant for any $g$) and, thus, the choice  $g(M_0)\simeq 1$
  appears to be a reasonable compromise. Note that in the
two-dimensional NLSM, a coupling of order $g(M_0) \simeq 1$ may be naively
compared  with a relatively strong four-dimensional QCD coupling
$\alpha_S\sim 1$, which is  well within the nonperturbative $T \sim T_c$
QCD regime.

\subsection{The $T=0$ results}

We start by considering the optimization solutions at $T=0$ and $N=4$. 

\begin{figure}[htb!]
\includegraphics[scale=0.5]{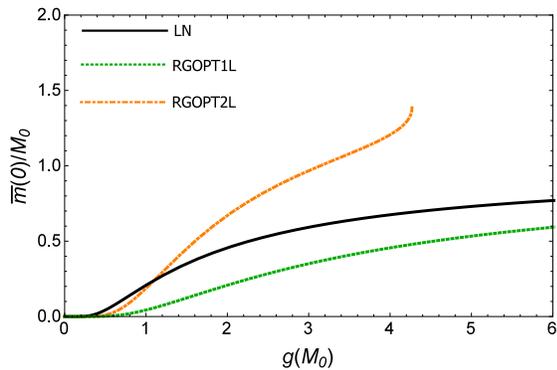}
\caption{\label{fig2} Normalized zero temperature optimized masses,
  ${\bar m}(0)/M_0$, as a function of $g(M_0)$, for  $N=4$ and at the
  central reference scale $\alpha=1$, in the RGOPT at one- and
  two-loops and in the LN approximations.}
\end{figure}

In {}Fig.~\ref{fig2} we compare for $N=4$ the optimized mass ${\bar m}(0)$
obtained with the one-loop RGOPT solution from (\ref{pms}), given
explicitly by Eq.~(\ref{mass1L}) in the $T\to 0$ limit, with the
similar $T\to 0$ two-loop RG solution
of $f_{full \, RG}^{(2L)}$ given by Eq.~(\ref{RGfull}),  and the LN mass,
as functions of the renormalized coupling at the central reference
scale $\alpha=1$. (The $T=0$ mass gap for the SPT is not
  shown as it is trivially vanishing from Eq.~(\ref{gapspt})).  The
results in {}Fig.~\ref{fig2} show that at one-loop order the RGOPT has
real solutions for all values of $g$. In contrast, the two-loop RGOPT
mass, using $f_{full \, RG}^{(2L)}$, Eq.~(\ref{RGfull}), becomes complex
beyond a rather high coupling value, for $N=4$, $g(M_0)\approx 4.27$, which is a
value high enough  for our purposes. This $g$ value, beyond which the
RG solution is complex, slightly decreases as $N$ increases, but for
$N=3$ one recovers a real solution for any $g$. It can be
verified that Eq.~(\ref{RGfull}) gives actually two branch
solutions. We select unambiguously the one which correctly reproduces
the SB result as the physical solution  in all  subsequent
evaluations. That is, as already mentioned concerning the other
solution from the mass optimization, Eq.~(\ref{pms2loop}), our
criterion to select $\bar m$ is to choose the root which reproduces
the perturbative results for small $g$. The other nonphysical
solution, not shown  in {}Fig.~\ref{fig2}, has an {\em anti-AF}
behavior, similarly to the other nonphysical solution of
Eq.~(\ref{pms2loop}), which is given by Eq.~(\ref{nonAF}).  (NB the
(real part of the) physical branch solution is not plotted beyond the
coupling value  where it starts to be complex, that is why it appears
to end abruptly).

\subsection{The $T\ne 0$ results}

\begin{figure}[htb!]
\includegraphics[scale=0.8]{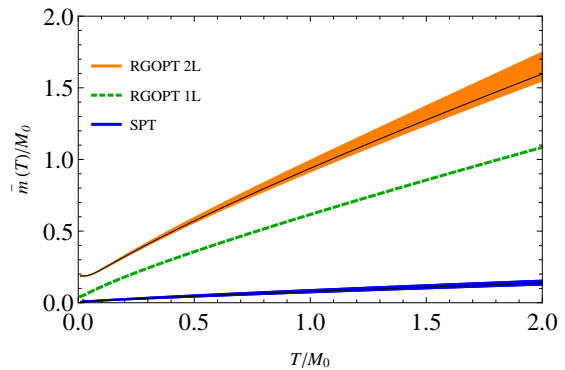}
\caption{\label{fig3} Thermal masses as a function of the temperature
  (both quantities normalized by the reference scale $M_0$)   for
  scale variations $0.5\leq \alpha \leq 2$, $N=4$ and
  $g(M_0)=g_{LN}(M_0)/2=1$. We compare the results for the  RGOPT
  (one- and two-loop cases) and the SPT.  (NB for this coupling choice
  the LN thermal mass is identical to the RGOPT one-loop one).  Within
  the two-loop RGOPT and SPT, the shaded bands  have the lower edge
  for $\alpha=0.5$ and the upper edge for $\alpha=2$. The thin line
  inside the shaded bands  is for $\alpha=1$.}
\end{figure}

When considering the finite temperature case,  we show in
{}Fig.~\ref{fig3} the one-loop RGOPT mass gap from
Eq.~(\ref{gap1}), the two-loop similar result from  Eq.~(\ref{RGfull}),
which now are functions of the temperature, fixing
$g(M_0)=1$ and varying the scale $M=\alpha M_0$, with $0.5 \leq \alpha
\leq 2$, corresponding to the shaded bands in {}Fig.~\ref{fig3}.   We
then see from {}Fig.~\ref{fig3} that the one-loop RGOPT mass is
exactly scale independent, as it was anticipated from its expression,
given by Eq.(\ref{gap1}) in the previous section. This is the case because, 
by construction, it satisfies both the RG
and the OPT equations simultaneously, Eqs.~(\ref {RGr}) and (\ref
{pms}), respectively, which lead to Eq.~(\ref{gap1}).  The two-loop
RGOPT  and SPT appear, however, not scale invariant. 
The reasons for this are as follows. 
First, the  residual scale
dependence of the SPT mass is not surprising,  since its construction
lacks RG invariance from the beginning, as already explained
previously.  Here this is very screened by the smallness of the perturbative mass Eq.~(\ref{mDB}) 
used for SPT.
While the two-loop RGOPT mass residual scale
dependence has a different origin: because the (perturbatively RG-invariant by construction)
mass obtained from any of the possible defining Eqs.(\ref{pms2loop})
or (\ref{RGfull}),  for arbitrary temperature, cannot match exactly
the running of the coupling Eq.~(\ref{run2}), dictated by the purely
perturbative behavior at zero-temperature.  Note that the scale
  dependence indeed increases for increasing $T/M_0$ in
  {}Fig.~\ref{fig3} (but this is artificially enhanced by showing $\bar m \sim T$ rather
  than $\bar m/T$, which rather decreases with $T$). The two-loop RGOPT scale dependence appears
  much larger than the SPT one, but this is essentially due to the
  intrinsically much larger $\bar m/T$ values
  in the RGOPT than in the SPT case, therefore within the
  $\bar m/T >1$ regime for intermediate $T/M_0$, where the implicitly high-$T$ regime justifying
  the   use of the perturbative running (\ref{run2}) is no longer
  quite valid.  
  However, as we will see below, this sizable scale
dependence of the RGOPT mass gets largely damped within  the RGOPT
pressure, giving an overall scale dependence much more moderate than
for the SPT pressure. 

\begin{figure}[htb!]
\includegraphics[scale=0.8]{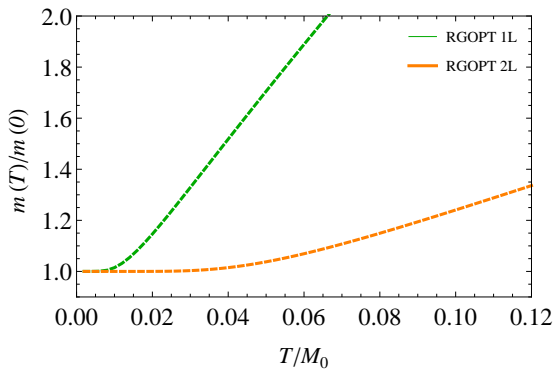}
\caption{\label{fig4} The normalized thermal optimized masses, ${\bar
    m}(T)/{\bar m}(0)$, as a function of the temperature $T$
  (normalized by $M_0$)   for $N=4$, $g(M_0)=1 = g_{LN}(M_0)/2$, and
  at the central scale choice $\alpha=1$, in the RGOPT  at one- and
  two-loop cases.  (NB for this coupling choice the LN thermal mass is
  identical to the RGOPT one-loop one).}
\end{figure}

The RGOPT mass ${\bar m}(T)$  clearly starts from a nonzero value at
$T=0$, since the mass gap solution is nontrivial at $T=0$ (Eq.~(\ref{mass1L})), then bends
and reaches, as expected by using basic dimensional arguments, a
straight line for large temperatures, where it behaves perturbatively as $\bar m\sim g T$.  As observed in
Ref.~\cite{giacosa},  this behavior is reminiscent of that of the
gluon mass in the deconfined phase of Yang-Mills
theories~\cite{YM1,YM2,YM3}, where, at high-$T$, the gluon mass can be
parametrized by $T/\ln T$.  The bending of the thermal masses can be
better appreciated in {}Fig.~\ref{fig4}, which shows that the changing
of behavior occurs at rather low temperatures.  

One should recall, as already emphasized, that $\bar m$ is only used
at intermediate steps and does not represent a direct physical
observable (see Refs.~\cite {jlprd,jlprl} for further discussions on
this issue). In this framework, physical quantities, like the pressure,
are  obtained upon substituting into these $\bar m(g)$, which
carries the nonperturbative coupling dependence.  Next, we compare the
results for the pressure obtained from the different schemes
considered, namely, the RGOPT, PT, SPT and LN.

\begin{figure}[htb!]
\includegraphics[scale=0.6]{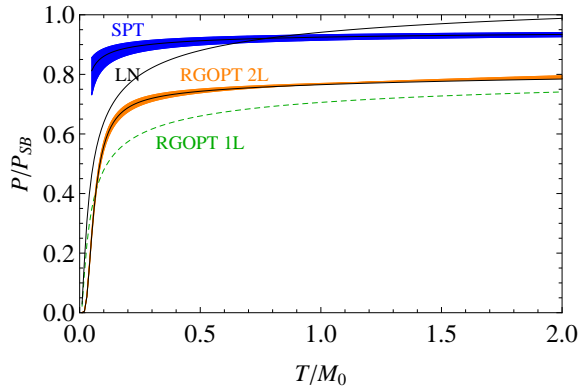}
\caption{\label{fig5} $P/P_{\rm SB}$ as function of the temperature
  $T$ (normalized by $M_0$) for $N=4$ and $g(M_0)=1 = g_{LN}(M_0)/2$,
  with scale variation $0.5\leq \alpha \leq 2$.  Within the two-loop
  RGOPT and SPT, the shaded bands  have the lower edge for
  $\alpha=0.5$ and the upper edge for $\alpha=2$. The thin line inside
  the shaded bands  is for $\alpha=1$.  (NB the line thickness of
  one-loop RGOPT and $LN$ is only for visibility at the figure scale
  and does not correspond to any actual scale dependence since those
  approximations are {\em exactly} scale invariant.)}
\end{figure}

In {}Fig.~\ref{fig5} we show the (subtracted) pressure, $P = P(T)-P(0)$,
normalized by $P_{\rm SB}$, for the scale variations $M=\alpha M_0$,
$0.5 \leq \alpha \leq 2$ and $N=4$.  It illustrates how the one-loop
RGOPT pressure is exactly scale invariant, while the two-loop result
displays a (small) residual scale dependence for the reasons already
discussed above when considering the mass and 
concerning the interpretation of the results shown in
{}Fig.~\ref{fig3}.  Note that despite the fact that the optimum mass
${\bar m}(T)$ has a non negligible scale dependence for the RGOPT
two-loop case, even compared to SPT, 
the RGOPT pressure itself exhibits a substantially smaller scale
dependence than the corresponding SPT approximation, at moderate and low $T/M_0$ values, 
as can be seen on {}Fig.~(\ref{fig5}).

While the improvement as compared with
 SPT  may appear not very spectacular at the two-loop order here illustrated, the
 important feature is that the RGOPT construction is expected on
 general grounds to  systematically improve the perturbative scale
 dependence at higher orders, as explained in
 Refs.~\cite{jlprl,jlprd}. Being built on perturbative RG
 invariance at order $k$ for {\em arbitrary} $m$, the mass gap
 exhibits a remnant perturbative scale dependence as $m(M) \sim g\,T [1+ \cdots
   +{\cal O}(g^{k+1}\ln M) ]$, such that the (dominant) scale
 dependence within the vacuum energy (pressure), coming from the leading
 term $\sim m^2/g$, should be of perturbative order $g^{k+2}$. 
 This feature can easily be checked explicitly in the above two-loop case:
 taking the high-$T$ perturbative expression of the pressure, Eq.~(\ref{rg2HT}),
 replacing $g$ there by its two-loop running, Eq.~(\ref{run2}), and tracing only
 the $M$ scale dependence, after a straighforward expansion one can check that it appears 
 first at perturbative order $g^3$, thus formally four-loop order: 
 \be
 \frac{ P^{\rm RGOPT}_{2L}}{P_{SB}} \simeq 1-\frac{3}{2}\,b_0 g(M_0)+{\cal O}(g^2(M_0))+{\cal O}(g^3\,\ln M).
 \label{remng3}
 \ee
 Of course the scale-dependence of the two-loop RGOPT pressure illustrated in Fig~(\ref{fig5}) reflects
 more than this naive perturbative behavior, since the exact pressure was used to describe correctly 
 the low-$T$ regime, where $\bar m$ has no longer a simple perturbative expansion expression. 
 In contrast, a completelly different behavior happens for SPT, or similarly the
 HTLpt. In analogy with the scalar model~\cite{jlprl},  in the NLSM
 the unmatched leading order remnant terms, $\sim m^2\ln M$ from
 Eq.~(\ref{remnant}), remain partly screened up to two-loop, since
 perturbatively $\bar m^2_{spt} \sim {\cal O}(g^2)$, but those
 uncancelled terms unavoidably resurface at the perturbative three-loop
 order $g^2$. Apart from improving the residual scale dependence, the very different properties
 implied by Eq.~(\ref{exponent}) and the $m^2/g$ term in Eqs~(\ref{P1L}), (\ref{rgopt2L}) 
 also explain the very different 
 shape (and lower values) of the RGOPT one- and two-loop pressures  as compared with SPT, 
 which also does not include $P(T=0)$. 

One can also
guess from {}Fig.~\ref{fig5} that the LN pressure overestimates the SB
limit, clearly from the changing $N-1\to N$ overall factor implicit in
the LN approximation, which results in a difference by a factor $4/3$
for the pressures when $N=4$. Both the one-  and two-loop RGOPT
pressures reach (very slowly, logarithmically with $T/M_0$) the SB
limit, that one cannot see on the scale of the figure.

\subsection{High-$T$ approximation and comparison with standard PT}
\label{sec6PT}

\begin{figure}[h]
\includegraphics[scale=0.8]{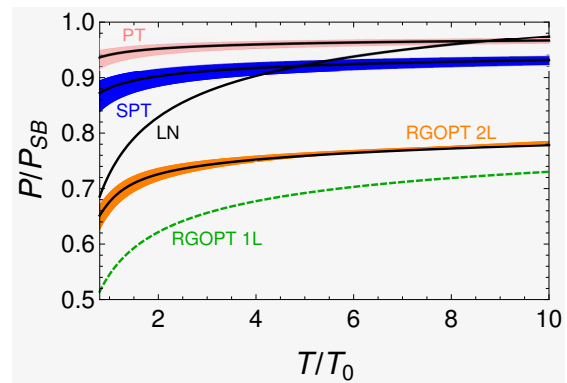}
\caption{\label{fig6} $P/P_{SB}(T/T_0)$ in high-$T$ approximation for $N=4$,
  $g(M_0=2\pi T_0)=1$, and scale variation $1/2 < \alpha<2$.
  Pink range: PT; blue range: SPT; black: LN;  orange range: two-loop RGOPT; 
  green dashed line: one-loop RGOPT.}
\end{figure}

Let us now illustrate the high-$T$ approximation, still for $N=4$, and
a scale variation with a factor $1/2 < \alpha < 2$. In
{}Fig.~\ref{fig6} we show $P/P_{SB}(T/T_0)$ for a fixed
  reference coupling, $g(2\pi T_0)=1$,  for the one and two-loop RGOPT
  cases in high-$T$ expansion approximation, Eqs.~(\ref{optHTsimp})
  and (\ref{rg2HT}), as compared with the standard PT result, given by
  Eq.~(\ref{Ppt}), and with two-loop SPT result
  Eq.~(\ref{Psptht}). (NB lattice results are not available in this
  high-$T$ regime). This  clearly displays again the exact scale
  invariance of the one-loop RGOPT pressure, while the two-loop RGOPT
  result has a moderate remnant scale dependence in comparison to the
  slightly more sizable SPT and standard PT scale dependences (for high $T$).  
  Notice that the scale dependence of SPT is somewhat more important than the standard
  PT one. Concerning the RGOPT, these results are just a straightforward
  restriction to the high-$T$ regime of the more complete arbitrary
  $T$-dependent features illustrated in {}Fig.~\ref{fig5}, except that
  here the unshown low $T/T_0 \lesssim 1$ behavior (which for $N=4$ and
  $g(2\pi T_0)=1$ corresponds roughly to $\bar m/T \gtrsim 1$ for the
  central scale $\alpha=1$ choice, see Eq.~(\ref{mbarHT})) is,
  therefore, not valid due to the intrinsic limitation of
  the high-$T$ expansion. This also explains why the RGOPT scale dependence improvement
  does not appear spectacular in the low $T$-regime, as compared with PT and SPT, while
  it was more drastic for the exact $T-$dependence on Fig.\ref{fig5}. 
  The fact that the one- and two-loop RGOPT
  pressure are still different for relatively large $T/T_0$, i.e.,  small
  $g(T/T_0)$, is clear from the two-loop extra terms comparing Eq.~(\ref{optHTsimp}) with
  Eq.~(\ref{rg2HT}). It is also clear from their latter analytical expressions that both the
one- and two-loop RGOPT pressures tend logarithmically towards the SB limit
for $T/T_0\to\infty$  (even if not obvious  from {}Fig.~\ref{fig6}), while the PT  and SPT pressures reach this limit
more rapidly.   

As explained above in Subsec.~\ref{secPT}, this visible difference  as
a function of $g$ of the one- or two-loop RGOPT pressures as compared
to the PT results, is actually perturbatively consistent for $g\to
0$. The correct matching appearing once considering the RGOPT in terms of the
physical input, which is like replacing the pole mass Eq.~(\ref{mDB})
within  Eq.~(\ref{optHTsimp}), which then gives the same perturbative
expansion, Eq.~(\ref{Ppt}).  Or equivalently, if expressing both
  the PT Eq.~(\ref{Ppt}) and RGOPT pressures as a function of the
  mass $m_{B}/T$ rather than the coupling, they have the very same
  slope for small $m/T$, see Eq.~(\ref{optHTsimp}), while the RGOPT
  pressure departs at larger $m/T$ from the PT one Eq.~(\ref{Ppt})
  due to the exact $m/T$ dependence in Eq.~(\ref{P1Lfin}). 

  The main interest in
  {}Fig.~\ref{fig6} is that it compares (qualitatively) more directly
  with the results of QCD  HTLpt~\cite{htlpt1,HTLPT3loop}, where at
  two-loop order and beyond, due to the quite involved gauge-invariant HTL framework, 
  only the high-$T$ expansion approximation has been
  worked out for the QCD HTLpt.  In this respect, it is
  instructive to compare our results in  {}Fig.~\ref{fig6} with
  those obtained, e.g., in Ref.~\cite{HTLPT3loop} and shown in
  {}Figs.~7-8 in that reference.  In particular, we observe that the
  shape and behavior of the SPT pressure is quite similar with the one- or two-loop
  HTLpt QCD pressures,  not departing much from the SB limit even for $T/T_0\sim 1$.
  However the HTLpt results have a very different behavior at three-loops~\cite{HTLPT3loop},
  departing very much from the SB limit at moderate and low $T$ values, and showing
  good agreement with lattice data for the central renormalization scale choice. 
  
  In contrast, the one- and two-loop RGOPT
  pressures appear to have a different, more nonperturbative behavior,
  in the sense that the RGOPT pressures show a more rapid bending to
  lower values for decreasing $T/T_0$ values, similarly to the 3-loop HTLpt results.  This behavior is also 
  roughly more qualitatively comparable to the lattice QCD simulation
  results for the pressure~\cite{karsch_glue,karschQCD}.   Therefore, we anticipate
  that a RGOPT application to QCD HTLpt is likely to give similar
  features as the present NLSM RGOPT pressure results that we have
  just obtained.  Indeed, we will illustrate below 
  the rather good agreement of the RGOPT NLSM pressure with lattice simulations~\cite{giacosa} for $N=3$.

\subsection{Comparison with next-to-leading $1/N$ expansion}
\label{sec61/N}

\begin{figure}[h!]
\includegraphics[scale=0.8]{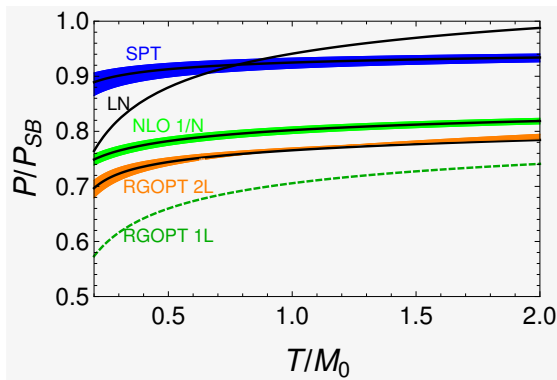}
\caption{\label{fig7}   $P/P_{\rm SB}(T/M_0)$
  in high-$T$ approximation for $N=4$, $g(M_0)=1=g_{LN}(M_0)/2$ and scale variation $1/2 < \alpha<2$. 
  Same captions as in Fig. \ref{fig6}, with added comparison to $1/N$ NLO pressure (green range).}
\end{figure}

{}Next, since the RGOPT incorporates finite $N$, it is of interest
to also compare our results with the ones obtained from the
nonperturbative $1/N$ expansion at NLO~\cite{warringa}. Since the $1/N$ NLO exact solution in
Ref.~\cite{warringa} is rather involved, to make the
comparison simpler  we consider the $1/N$ NLO pressure in the 
high-$T$ approximation only. It reads~\cite{warringa}

\begin{equation}
P_{1/Nnlo} \approx T^2 \frac{\pi}{6}(N-1) -
\frac{T}{4}(N-2)m_{1/Nnlo}  ,
\label{P1/N}
\end{equation}
where

\bea 
m_{1/Nnlo}&\approx& T \pi \left [\left
  (\frac{2\pi}{g_{LN}}- \frac{\ln 4}{N} \right ) \left ( 1+ \frac{2}{N} \right )
  \right .  \nn \\ &-& \left .  \gamma_E -\ln \left ( \frac{M}{4 \pi
    T} \right ) \right]^{-1} ,  
\label{m1/N}    
\eea
 where the (rescaled) coupling has the same meaning as the LN one, Eq.~(\ref{LNresc}).
First, notice that already the one-loop RGOPT mass gap, Eq.~(\ref{mbarHT}), upon expanding it in $1/N$, 
has a quite similar expression as (\ref{m1/N}), only missing the $\ln 4/N$ term. 
In {}Fig.~\ref {fig7} we show the pressure, all evaluated in the high-$T$
limit and obtained with the different approximations,  with the scale dependence illustrated as 
previously, for $1/2 < \alpha <2$.  It clearly displays how, from one to  
two-loop, the RGOPT pressure appears to converge rather well 
to the NLO $1/N$ result.  Note also that the nonperturbative NLO $1/N$ pressure exhibits a residual, 
though moderate, scale dependence: this comes from the fact that the exact NLO $1/N$ running coupling~\cite{warringa}, 
which is given by Eq.~(\ref{run1}) upon rescaling the coupling from Eq.~(\ref{LNresc}), 
does not perfectly match the scale dependence of Eq.~(\ref{m1/N}). 
This is analogous to the origin of the residual scale dependence of the two-loop RGOPT pressure,
already explained above, which indeed appears in Fig.~\ref {fig7} to be of a similar range as the NLO $1/N$ scale dependence. 
\subsection{Further improving residual scale dependence}
\label{sec6geff}

 We have previously explained why, despite the explicit perturbatively
 RG invariant RGOPT construction, there is a moderate residual scale
 dependence within the two-loop results. 
Now, given that we restricted our analysis at the two-loop order,
one may wonder generally if it is possible to further improve the
RGOPT scale invariance at this order.  As mentioned in the
introductory part of Sec.~\ref{sec5},  if we could obtain a
simultaneous solution of both Eqs.~(\ref{pms2loop}) and
(\ref{rg2loop}), therefore getting rid of using the perturbative
(massless) running from Eq.~(\ref{run2}), we could intuitively expect
a further reduced, minimal scale dependence (but still not expecting
perfect scale invariance at a given finite order of the RGOPT modified
expansion, which obviously remains as an approximation to the truly
nonperturbative result).  But since in the NLSM  such nontrivial
solution does not exist at two-loop order, an interesting question
that we address now is whether  the scale dependence may be further
improved nevertheless, by using some alternative procedure.

In fact it is only the combination of the {\em exact} two-loop mass
optimization prescription (MOP) Eq.~(\ref{pms2loop}) with the exact RG
Eq.~(\ref{rg2loop}), which leads to the trivial solution $g=0$,
Eq.~(\ref{rg+opt}).  Now, since RGOPT is still a perturbatively
constructed approximation, it is perhaps too contrived to require such
exact solutions at two-loop order. Thus, one possible trick is to bypass
the actual triviality of the combined solution by simply approximating one of
the two equations. Typically, the simplest procedure is to keep  the
RG Eq.~(\ref{rg2loop}) as exact, giving the mass gap $m(g(T))$ just as
previously,  while considering the other exact physical solution of
Eq.~(\ref{pms2loop}), given by Eq.~(\ref{gap1}), as defining an {\em
  effective} temperature-dependent coupling $g(T)$, but approximating
the latter by truncating the thermal mass to its purely perturbative
first order high-$T$ expression, Eq.~(\ref{mbarHT2}). This
gives the result
\bea 
& b_0\,g_{eff}(m,M) = \left[ \ln \frac{M}{m} -2
  J_1(\frac{m}{T}) \right]^{-1} \nn \\  & \simeq  \left\{ \ln
\left[\frac{2M}{(N-2)g\,T}\right] -2 J_1(\frac{(N-2)g}{2} )
\right\}^{-1}  .
\label{geffrgopt}
\eea 
There is one minor subtlety at this stage: while the standard
perturbative running, e.g., Eq.~(\ref{run1}) at one-loop order, is
calibrated such that the central value $\alpha=1$  corresponds exactly
to $M=2\pi T$, this has no reason to be the case  for the running
coupling (\ref{geffrgopt}), having a more nonperturbative 
dependence. Thus, to get a sensible comparison of scale dependence,
namely for identical central coupling values $g(M_0)$, we have to
match $M_0$ such that $g_{eff}(M=\alpha M_0)\equiv g(M_0)$ for
$\alpha=1$, which is obtained for an appropriate central value
$\alpha$ upon setting $M\equiv 2\pi \alpha$ in
Eq.~(\ref{geffrgopt}). More precisely, for $g(M_0)=1$  this matching
happens for $\alpha\simeq 2 e^{-\gamma_E}/2\simeq 1.12$, a very
moderate shift of renormalization scale calibration. 

\begin{figure}[h]
\includegraphics[scale=0.8]{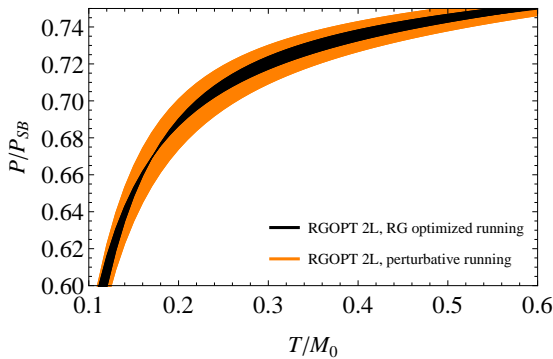}
\caption{\label{fig8} $P/P_{SB}$ (zoomed in {}Fig.~\ref{fig5})
  for the two-loop RGOPT pressure, comparing using the perturbative
  two-loop running coupling Eq.~(\ref{run2}),  or the truncated RG
  optimized running coupling Eq.~(\ref{geffrgopt}).  For the latter, for $T/M_0\gtrsim 0.17$ 
  the lower border
  of the range corresponds to the lower scale choice $\alpha=0.5$,
  and the upper border corresponds to the higher scale choice
  $\alpha=2$.  While for  $T/M_0\lesssim 0.17$ it is the opposite.}
\end{figure}

Now, by combining the approximate solution Eq.~(\ref{geffrgopt}) with
the Eq.~(\ref{rg2loop}), we do obtain a nontrivial coupled $\{\bar g,
\bar m\}$ solution, and Eq.~(\ref{geffrgopt}) has sound properties of
an effective coupling, being consistent (at least at one-loop order)
with the standard running coupling (\ref{run1}).  Morever, we stress 
that Eq.~(\ref{geffrgopt}) is not  put by
hand, but it is a direct consequence of Eq.~(\ref{gap1}), at a minimal
extra work cost, since one already knows all the  ingredients from the
above calculations at this stage. Using 
Eq.~(\ref{geffrgopt}), one can examine the resulting scale dependence
that follows from this alternative procedure. This is illustrated for
the pressure, as compared with the previous results using the purely
perturbative running (\ref{run2}),  in {}Fig. \ref{fig8}.  One
observes a definite further improvement, by a factor two roughly,
specially in the  intermediate $T/M_0$ zoomed in region, where the
previous scale dependence was the largest.

We have tried to pursue this construction by truncating the RGOPT
thermal mass $m$ at the next two-loop order,  but the scale dependence
does not further improve, rather worsening.  This ``incompressible",
minimal residual scale dependence beyond one-loop order  is
actually expected from general perturbative arguments, 
since one cannot expect to have exact scale invariance at two-loop
order or beyond.  Note, moreover, that if one would push farther such
approximations of Eq.~(\ref{gap1}), including more and more
perturbative terms in the thermal mass expansion, one would
progressively see the combined solution unavoidably reach the exact
trivial one, Eq.~(\ref{rg+opt}), $g_{eff}\to 0$.  Therefore,
Eq.~(\ref{geffrgopt}) appears to be a relatively simple and optimal
effective prescription to minimize the scale dependence.  Remark
  also that Eq.~(\ref{geffrgopt}) has essentially a one-loop running
  form, when comparing with Eq.~(\ref{run1}), but the arbitrary
  $T$-dependence that it involves allows to
  consider a more nonperturbative coupling $g$ regime, as we will
  illustrate next, when comparing with lattice results. 

Now, as we have already emphasized before, since at two-loop RGOPT
order the correct AF branch mass gap solution behaves perturbatively 
essentially like the one-loop solution, Eq.~(\ref{gap1}), it is not too
suprising that Eq.~(\ref{geffrgopt}) appears as an optimal
running. But we must emphasize that this feature is very peculiar to
the NLSM model at two-loop order, as already explained
previously, with the factorized form of the two-loop mass gap solution
(\ref{pms2loop}) due to the specific renormalization properties of the NLSM vacuum energy. 
More generally for other models, the running coupling at a $n$-loop order should be
optimal for the RGOPT at $n$-loop order.  

\subsection{Comparison with lattice simulations}
\label{sec6latt}

\begin{figure}[htb!]
\includegraphics[scale=0.6]{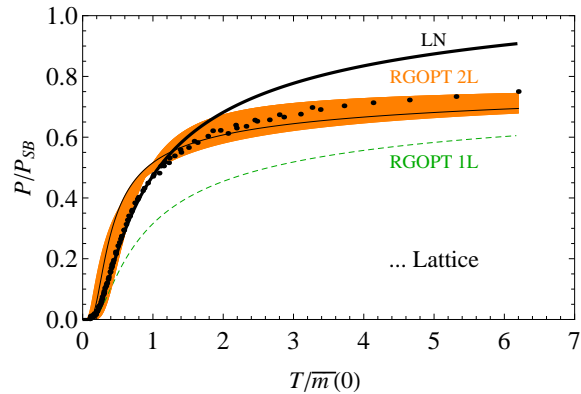}
\caption{\label{fig9} $P/P_{\rm SB}$ as a function of the
  temperature $T$ (normalized by the $T=0$ mass gap $\bar m(0)$) for $N=3$: LN, one- and
  two-loop RGOPT for  $g(M_0)=g_{LN}(M_0)=2\pi$ and scale variation $1/2 \leq
  \alpha \leq 2$, when using RG optimized running  given by
  Eq.~(\ref{geffrgopt}), compared with lattice simulations (taken from
  Ref.~\cite{giacosa}).  NB lattice data have been conveniently
  rescaled on vertical axis from $P/T^2$ in Ref~\cite{giacosa} to
  $P/P_{SB}$ (i.e., for $N=3$ this corresponds to a scaling factor of
  $\pi/3$).}
\end{figure}

We will now compare the RGOPT results with lattice simulation ones.
 Apart from the early work in \cite{lattice_early}, to the best of our knowledge the only
available lattice thermodynamics simulation of the NLSM is the one of 
Ref.~\cite{giacosa}, which was performed for
$N=3$~\footnote{We thank the authors of  Ref.~\cite
    {giacosa} for providing us their lattice data for the pressure.}.  
    To complete this comparison, we need  a priori to fix an
appropriate coupling value at some scale $M_0$, recalling that the simulation in
Ref.~\cite{giacosa} was performed at relatively strong lattice
coupling values. As required in lattice simulations, the authors consider a
sequence of different (bare) lattice couplings for different $T/M_0$ ranges,
in order to best control the approach to the continuum. 
Our analytical result is evidently aimed for fixing a $g(M_0)$ input
choice (its running with the scale being determined from RG properties). 
Moreover the RGOPT approximation effective coupling, in the $\ms$ scheme, 
has no reasons to coincide with the lattice coupling definition. 
As already explained above, the combined two-loop solution for the RG Eq.~(\ref{rg2loop}) and the MOP
Eq.~(\ref{pms2loop}), if it would exist at two-loop order, would determine besides the optimal mass similarly an 
optimal $T$-dependent coupling $\bar g(M,T)$, thus giving a compelling choice for
comparing with lattice results. 
In absence of such optimal two-loop coupling for
the NLSM two-loop results, there is, however, one other remarkable
coupling value $g(M_0)$ (at two-loop order), namely such that $\ln \bar
m(0)/M_0 =0$ exactly (i.e. such that the zero-temperature mass gap
coincides with the scale $M_0$, with no further corrections).  This
happens for $g(M_0)=2\pi$ (a value coincidentally analogous
to the one obtained in the GN model~\cite{JLGN} for the exact
optimal coupling).  It thus appears
to be a sensible input choice to compare with
lattice, as it is also in the nonperturbative regime.  In
{}Fig.~\ref{fig9} we thus compare the one- and two-loop RGOPT
and the LN pressure for $g(M_0)=2\pi$  with the lattice data for
$N=3$,  as function of the temperature, now normalized  by 
the $T=0$ mass gap $\bar m(0)$, consistently with the lattice results normalization\cite{giacosa}. \\  
 Our LN pressure  exactly coincides analytically and numerically with the one in Ref.~\cite{giacosa}). 
The one-loop RGOPT and LN pressure are exactly scale
independent for any $g$, as already pointed out before 
  (remarking again that the only difference between one-loop RGOPT
  and LN is the $N-1\to N$ overall factor). It is also worth noting that, 
  once expressed as $P(T/\bar m(0))$, 
both the one-loop RGOPT and LN pressures are actually {\em independent} of the input coupling value $g(M_0)$: 
the reason for this is that the one-loop pressure
Eq.~(\ref{P1Lfin}) (and therefore also its large-$N$ limit) do not depend explicitly on $g$, while 
the $g(M)$ dependence is entirely absorbed within $\bar m(0)$ from the mass gap Eq.~(\ref{gap1}).
(Accordingly the one-loop and LN pressures in Fig~\ref{fig9} do not get closer nor farther from the lattice data for 
other $g(M_0)$ choices). \\
But this remarkable feature holds only approximately at two-loop order,
due to the no longer matched $g(M)$-dependence between the pressure Eq.~(\ref{rgopt2L}) and 
the two-loop mass gap equation.
Moreover this mismatch is strongly enhanced in the peculiar $N=3$ limit in the NLSM, due to the vanishing of 
the two-loop $g$-dependence in 
the pressure Eq.~(\ref{rgopt2L}). Consequently, for $N=3$ the two-loop RGOPT pressure 
happens to be accidentally much sensitive to the $g(M_0)$ input choice.
An additional drawback is that for such large reference coupling $g(M_0)\sim 2\pi$, 
the unavoidable residual two-loop RGOPT scale dependence,  which
had remained very moderate for $g(M_0)\lesssim 2$, is now substantially
enhanced. We thus use the further improved alternative running
coupling, determined by Eq.~(\ref{geffrgopt}), but the residual scale
variation appears quite sizable. Nevertheless,
  when we  compare the RGOPT results with the scale dependence of
  other approximations, such as the PT or SPT for $N\neq 3$ (which for
  such large coupling values is well beyond any reasonable
  variation), the much better performance of the RGOPT is quite
  visible. (Remark that we do not illustrate PT and SPT here for $N=3$,
  since at two-loop order the SPT mass gap solution
  (\ref{mDB}) trivially vanishes  for $N=3$, consequently giving in Eq.~(\ref{Pspt}), (\ref{Psptht}) a trivial (constant) 
 pressure equal to the SB limit for any temperature).\\
 Within this sizeable scale variation uncertainty, the two-loop
RGOPT pressure shows a reasonably good agreement with lattice results  for $T \gtrsim \bar m(0)$,
but not so good, at least for the central renormalization scale, for low $0.2 \lesssim T/\bar m(0) \lesssim 1$ values.
 However, due to the above explained important sensitivity of the two-loop pressure to the input coupling for $N=3$,   
the value $g(M_0)=2\pi$ has to be considered more 
a reasonably good 'fit' of the lattice data from the two-loop results, coincidentally, than a genuine prediction of RGOPT. 
It is intriguing that this value has other independent motivations, 
but it is probably more correct to conclude that this reasonable agreement 
is largely coincidental.
   
  Another rather stricking result is the extremely good agreement of lattice data with LN for
  small to moderate $T/m(0) \lesssim 1$. 
  While for very large $T$ values the lattice (and also the two-loop RGOPT) pressures are
  consistent with the true SB limit of the model Eq.~(\ref{SB}), $P_{SB}\sim (N-1)$. This visible 
  difference between the low- and high-$T$ regimes of the NLSM model is an important nonperturbative crosscheck: 
  at sufficiently low $T$, nonperturbative results should reflect the unbroken $O(N)$ with actually $N$ degrees of freedom.
  In the RGOPT framework, similarly to the LN approximation,  as we have explained
  the constant vacuum energy piece $m^2/g$ (footprint of a $\sigma$ field term), plays a crucial
  role in obtaining a mass gap with these expected features of the low-$T$ nonperturbative NLSM properties. 
  While at asymptotically high-$T$ one reaches the free theory $g\to 0$ limit of the NLSM model, thus
  describing a gas of $N-1$ non-interacting pions, while the non-kinetic $m^2/g$ contribution becomes negligible.   
  The RGOPT two-loop results roughly exhibit this overall nonperturbative behavior from low- to high-$T$ regime (although not
  perfectly at very low temperatures).  

\subsection{The Trace anomaly}
\label{tranom}

It is also of interest to investigate the behavior of some other
thermodynamical quantities evaluated in the RGOPT scheme and how they
compare with the same quantities evaluated in the SPT and LN
approximations.  {}For example, the  interaction measure
\footnote{As a rather trivial remark, one notes that
this differs from the usual expression quoted in the literature simply because
here we are working in 1+1-dimensions, while in 3+1-dimensions the pressure
obviously appears multiplied by a factor 3. Likewise, the normalization here
is only $T^2$, instead of $T^4$ in 3+1-dimensions.} $\Delta=
({\cal E}- P)/T^2 \equiv T\partial_T[P(T)/T^2]$, 
which is the trace of the energy-momentum tensor
normalized by $T^2$. The  interaction measure  
can be readily obtained from the pressure
by using the definitions for the entropy density, 

\be 
S= \frac{d}{d\,T} P(\bar m(g,T,M),T,M),
\label{Sdef}
\ee 
and for   the energy density, ${\cal  E}= -P +S\, T$.

One subtlety rooted in the optimized perturbation, is that one should
evaluate the entropy  according to Eq.~(\ref{Sdef}), namely only {\em
  after } the mass gap $\bar m(g,T)$, which is explicitly
$T$-dependent, had been  replaced within the pressure expression,
which, thus, becomes only a function of $g(T/M),T/M$. While the
partial derivative calculated for fixed (arbitrary) $m$,  ignoring the
mass gap (which would give the correct result for the truly massive
theory), is generally different and would lead to physically
inconsistent results in the present case. Actually, at the one-loop
RGOPT order,   since the mass gap is defined by the constraint given
by (\ref{pms}) and that results in Eq.~(\ref{gap1}), the two
expressions coincide. This here is simply because in $dP(m(T),T)/dT
=\partial P/\partial T +(\partial m/\partial T) (\partial P/\partial
m) $ the last term vanishes by definition  due to
Eq.~(\ref{pms}). While for the two-loop RGOPT, our mass gap
prescription that guarantees real solutions  is given by the RG
Eq.~(\ref{RGfull}), which is obviously not equivalent to
Eq.~(\ref{pms}). Likewise, the SPT mass gap, since it can only be
real-defined at this order as the Debye mass (\ref{mDB}),  is
also very different from the optimized mass (\ref{pms}).  At the
one-loop RGOPT order, we easily obtain simple compact analytical
expressions for the entropy and trace anomaly, given, respectively, by

\be 
S^{\rm RGOPT}_{1L}= -T \frac{(N-1)}{\pi} \left[2 J_0(\bar x)+ \bar
  x^2 J_1(\bar x) \right],
\label{Srg1}
\ee 
and

\be 
\Delta^{\rm RGOPT}_{1L}(T)-\Delta(0) =
\frac{(N-1)}{4\pi} \left[\bar x^2(T)-\bar x^2(0)\right], 
\label{Drg1}
\ee 
where it is understood that $\bar x\equiv \bar m(T)/T$ is given by
the solution of  Eq.~(\ref{gap1})~\footnote{Note that despite the
  overall negative sign of expression (\ref{Srg1}), this entropy
  remains positive definite for any $x$, as expected.}.  Note that we
normalize $\Delta$ by subtracting its $T=0$ expression, consistently
with the pressure normalization, $\Delta(0)\equiv -2P(0)/T^2=(N-1) \bar
m^2(0)/(4\pi T^2)$.  The corresponding LN expressions are very similar to
Eq.~(\ref{Srg1}) and (\ref{Drg1}), being obtained by the appropriate
LN rescaling (\ref{LNresc}) and overall $N-1\to N$ usual changing.

The two-loop SPT interaction measure has also a 
 relatively simple expression, due to the SPT mass gap being the
perturbative pole mass (\ref{mDB}) at this order, with $T$-dependence such that $\bar m'_{spt}(T)= m_{spt}/T 
(1-b_0 g+{\cal O}(g^2))$. 
(note also that $\Delta_{spt}(T=0)$ trivially vanishes).
We obtain
\bea 
& & \Delta^{\rm SPT}_{2L}(T)  = 
\frac{g\,(N-1)}{16\pi^2} \bar x^2  \left[ 4\pi b_0 (1+2 \bar x^2 J_2(\bar x)) \right. \nonumber \\ 
& & \left. -(N-3) \left\{ 1-b_0 g (1+2 \bar x^2 J_2(\bar x)+\frac{3}{2} Y)\right\} Y  \right] ,
\label{Dspt}
\eea
where $\bar x_{spt} =\pi\gamma_0\,g(T)$ from (\ref{mDB}) and $Y$ was defined in Eq.~(\ref{Ydef}). Thus 
its main feature is that Eq.~(\ref{Dspt}) is substantially smaller, at high and moderate $T$ values, 
relative to e.g. (\ref{Drg1}), 
being suppressed by the 
small SPT mass gap $\sim x^2_{spt}$, and has an essentially monotonic dependence on $T$. \\
As concerns the two-loop RGOPT, the
exact analytical expressions of $S$ and $\Delta$  are rather involved
and not very telling, due to the more involved
nonperturbative mass gap $\bar m(g,T)$ expression, which we recall is obtained by solving
Eq.~(\ref{RGfull}). Thus, we refrain ourselves from writing down
explicitly its full expression and we will only present
its corresponding numerical result. As we will illustrate, however, its
shape as function of $T/M$ is to a good approximation roughly similar to the simple one-loop RGOPT in (\ref{Drg1}).

\begin{figure}[htb!]
\includegraphics[scale=0.6]{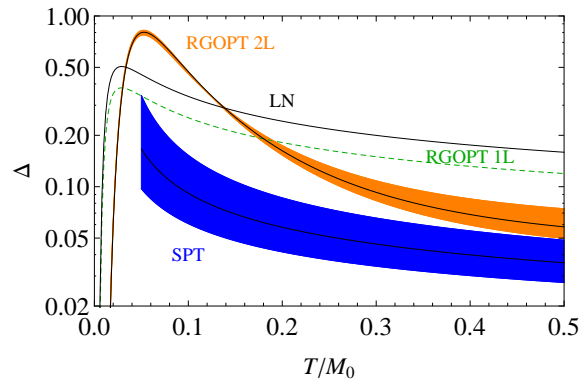}
\caption{\label{fig10} The interaction measure $\Delta$ as
  a function of the temperature $T$ (normalized by $M_0$)  for $N=4$
  and $g(M_0)=1 = g_{LN}(M_0)/2$, with scale variation  $0.5\leq
  \alpha \leq 2$, using the standard two-loop running coupling given
  by Eq.~\ref{run2}).  Within the two-loop RGOPT and SPT, the shaded
  bands  have the lower edge for $\alpha=0.5$ and the upper edge for
  $\alpha=2$. The thin line inside the shaded bands  is for
  $\alpha=1$. (Logarithmic scale is used).  }
\end{figure}

In  {}Fig.~\ref{fig10} we show the dependence of the interaction
measure as a function of the temperature in the one- and two-loop
RGOPT, two-loop SPT, and LN cases, for the same  choice of $N=4$ and
$g=1=g_{LN}/2$, as in the previous pressure plots shown in
{}Figs. \ref{fig2}-\ref{fig5}.

We notice from the RGOPT results shown in {}Fig.~\ref{fig10} how the
inflection before the peak  of $\Delta$ occurs approximately at the
temperature value where $m(T)$ bends (see {}Fig.~\ref {fig4}), which
is an interesting feature if one recalls that in QCD the inflection
occurs at $T_c$. It is worth to trace  more precisely 
the origin of the peak of the RGOPT interaction measures. First, note that 
$\bar x^2(T/M_0) = \bar m^2(T/M_0)/T^2$, as determined 
from Eq.~(\ref{gap1}), is monotonically decreasing. Thus the peak only originates from its interplay with the subtracted 
zero-temperature reduced (squared) mass gap $\bar x^2(0)\equiv \bar m^2(0)/T^2$ in Eq.~(\ref{Drg1}). It is easy 
to take the derivative with respect to $T$ of  (\ref{Drg1}), upon having first 
determined, from Eq.~(\ref{gap1}), that
\be
\bar m'(T) = 2 \bar x^3 \frac{J_2(\bar x)}{1+2\bar x^2 J_2(\bar x)} ,
\ee
to trace analytically that (\ref{Drg1}) has an inflection point, and then 
a maximum at a given $t_m\equiv T_m/M_0$ value, determined from the
solution of 
\be
\bar x^2(0) \equiv \frac{e^{-\frac{2}{b_0\,g}} }{t_{m}^2} = \frac{\bar x^2(t_{m})}{1+2\bar x^2(t_{m})\,J_2(\bar x(t_{m}))} .
\ee
{}For $N=4$, $g(M_0)=1$, it gives $t_{m} \simeq 0.029$, $\bar x(t_{m})\simeq 1.95$, which corresponds 
exactly to the peak position in {}Fig.~\ref{fig10}. (Note that the peak position is the same for the LN trace anomaly, for this
choice of couplings). One also easily derive the peak value as

\begin{eqnarray}
\Delta^{\rm RGOPT}_{1L}(t_{m})-\Delta(0) &=&
\frac{(N-1)}{2\pi}\frac{e^{-\frac{2}{b_0\,g}} }{t_{m}^2}\bar x^2(t_{m})
\nonumber \\
&\times& J_2(\bar x(t_{m}))
\end{eqnarray}
which gives $\simeq 0.38$, like is seen in {}Fig.~\ref{fig10}. For the two-loop RGOPT interaction measure, it is more difficult
to trace analytically but this quantity follows similar features, except that the bending of $\bar m$ is delayed to larger 
$T/M_0$ (see again {}Fig.~\ref {fig4}), so are the corresponding inflection point, and subsequent peak in {}Fig.~\ref{fig10}.
 Thus, although there is no phase transition in two dimensions, the trace anomaly has a nontrivial
structure with an inflection point followed by a peak, due to the occurence of a mass gap,
signaling the breakdown of scale invariance already at $T=0$, and the nontrivial $T$-dependence of this mass gap:
since $\bar m(0)$ in Eq.~(\ref{mass1L}) reflects dimensional transmutation, (\ref{Drg1}) (or similarly its two-loop generalization)
involve nonperturbative power contributions.
Note also that the RGOPT prediction is that
${\cal E}-P\equiv \Delta\,T^2$ grows quadratically with $T$ at high temperatures,  up to $\ln T$ terms, as easily established   
from the high-$T$ expression of (\ref{Drg1}) using $\bar x(T)$ from Eq.~(\ref{mbarHT}): $\Delta(T\gg M_0)\simeq 1/\ln (T/M_0)$. 
This behavior may be viewed as the two-dimensional analog, and  qualitatively compared with
four-dimensional Yang-Mills theory where 
a quadratic behavior has been found in the LQCD evaluations
performed in Ref.~\cite{karsch_glue}, and analytically supported by convincing arguments in \cite{quadraticDelta,RArriola}.  

Another interesting feature
shown by the results in {}Fig.~\ref{fig10}  is that the
two-loop RGOPT predicts that after the ``transition" (inflection) the
system interacts in a much stronger way than predicted by the LN and
the one-loop RGOPT, which have smaller thermal masses (see
{}Fig.~\ref{fig3}).   In this respect, it is again instructive to
compare our results for the interaction measure, given by
{}Fig.~\ref{fig10}, with  those obtained in Ref.~\cite{HTLPT3loop} and
shown in {}Fig.~12  in that reference.  While the leading-order (LO) and next-to-leading order (NLO)
HTLpt results show an essentially perturbative behavior, the three-loop (NNLO) 
HTLpt interaction measure is closer to the LQCD simulations~\cite{karsch_glue,karschQCD} (although not 
reproducing the lattice data peak). {}From this comparison
one can appreciate that the two-loop RGOPT interaction
measure has a shape similar to the one obtained in the LQCD
simulations.
In contrast the two-loop SPT interaction measure looks qualitatively more similar to 
the three-loop HTLpt results~\cite{HTLPT3loop},   
which we understand as originating from the accessible exact low $T$ dependence in the NLSM case.
It is a monotonic function of $T$ with 
no peak at any $T/M_0$ (as we also checked 
from Eq.~(\ref{Dspt}) and going further below the $T/M_0$ values shown in {}Fig.~\ref{fig10}).

\section{Conclusions }
\label{sec7}

We have applied the recently developed RGOPT nonperturbative framework
to investigate thermodynamical properties of the asymptotically free
$O(N)$ NLSM in two dimensions, and illustrate results for $N=3$ and $N=4$. Our application shows
how simple perturbative results can acquire a robust nonperturbative predictive power 
by combining  renormalization group properties with a variational criterion used to fix the (arbitrary)
``quasi-particle" RGOPT mass. 

In particular, a non-trivial scale invariant result was obtained by
considering the lowest order contribution to the pressure, which
represents a remarkable result if one considers that the whole
large-$N$ series can be readily reproduced upon taking the $N \to
\infty$ limit within the RGOPT.  In addition, at realistic finite $N$
values and high temperatures, the lowest order RGOPT pressure
converges to the correct Stefan-Boltzmann limit, while the LN result
overshoots it. Next, in accordance with other previous finite temperature
applications~\cite{jlprd,jlprl},  the NLO (two-loop) order RGOPT results display a very
mild residual scale dependence when compared to the standard SPT/OPT
results. 
The much reduced residual scale dependence is due to the explicitly RG invariant construction at all stages, as we recall:
first from retaining (or reintroducing, if absent) appropriate finite vacuum energy subtraction, Eq.~(\ref{skdef}),
to restore perturbative RG invariance of the vacuum energy of the model. Second, by maintaining RG invariance while
modifying the perturbative series with a generalized RG-dictated interpolation, Eq.~(\ref{delta}) with (\ref{exponent}).  
In contrast in related variational SPT or HTLpt approaches, the vacuum energy subtraction are omitted, and the
simpler linear interpolation is used. 
One should remark however that in thermal theories,
the omission of additional vacuum energy term is essentially innocuous at lowest order, 
since the thermal mass $m_T$ has itself a perturbative origin: here for the two-dimensional NLSM, $m_T\sim g T+{\cal O}(g^2)$, 
(see Eq.~(\ref{mDB})),   
such that, if uncancelled, the remnant term (\ref{remnant}) is formally of higher ${\cal O}(g^2)$ order.
These features are completely similar in four-dimensional models~\cite{jlprd}, where the vacuum energy involves $m^4$ terms,
while the thermal mass behaves for small $g$ as $m^2_T \sim g T$. 
Therefore the SPT or HTLpt formal lack of scale invariance at one-loop order is essentially ``screened'' by thermal masses, 
at least as long as $g$ takes perturbative values. But conversely 
it essentially explains why a more dramatic scale dependence is seen to resurface at higher orders, 
in particular at three-loop order in resummed HTLpt~\cite{HTLPT3loop}.  

We also obtain a reasonable agreement of the RGOPT pressure with known lattice results for $N=3$,
in the full temperature range,  with the expected nonperturbative behavior of the NLSM 
from low- to high-$T$ regime (but the agreement with lattice results is not quite good at low temperature).
However, this agreement is largely accidental at two-loops, coincidentally  
for a somewhat large input coupling choice $g(M_0)\simeq 2\pi$.
We remark that these rather odd properties of the two-loop NLSM RGOPT results essentially originate from 
the perturbative two-loop pressure contribution vanishing for $N=3$, see Eq.~(\ref{rgopt2L}), therefore inducing a severe mismatch 
in the good scale invariance properties otherwise verified for any other $N>3$. We can speculate an a priori 
much better behavior at three-loops for $N=3$, having simply determined from lower orders, using solely RG invariance properties, 
that the three-loop pressure contribution does not vanish for $N=3$, but a more precise investigation is well 
beyond our present scope.

The NLSM thermodynamical observables obtained from two-loop RGOPT display  a physical behavior that is more in line
with LQCD predictions for pure  Yang-Mills four-dimensional theories, as compared with the two-loop order SPT.   
Perhaps the most striking result, also in view of applications to QCD, is that 
the one- and two-loop RGOPT interaction measure $\Delta$ exhibit some characteristic nonperturbative
features somewhat similar to the QCD interaction measure.
Although as previously explained the underlying physics is very different since in two dimensions 
there is no phase transition, and the inflection and peak in the NLSM interaction measure reflect simply 
the broken scale invariance from a mass gap.
Yet the underlying mechanism appears here simpler but 
somewhat similar to the one advocated for QCD~\cite{quadraticDelta,RArriola,giacosaDelta},
in the sense that this peak originates from an interplay between thermal perturbative and nonperturbative
$T=0$ (power) contributions, present in RGOPT results. 
One may qualitatively compare these features 
with the HTLpt interaction measure, which is much closer to the lattice QCD results at 
three-loops~\cite{HTLPT3loop},  but does not show a transition peak.
While this is made possible in the present NLSM case due to the rather simple
structure of the model, giving an analytical handle to the full temperature dependence up to two-loop order. 
Apart from such possible technical limitations for a similar application to thermal QCD, the present 
NLSM results nevertheless confirm that the recently proposed RGOPT
approach stands as a robust analytical tool to treat renormalizable theories at extreme conditions. 

 Finally it would be of much interest to compare our NLSM thermodynamical results 
 with other lattice simulation results for other $N$ values,
  but unfortunately to our knowledge no such simulations at finite temperature are available up to now for $N>3$.

\section*{Acknowledgments}

MBP and ROR are partially supported by Conselho Nacional de
Desenvolvimento Cient\'{\i}fico e Tecnol\'{o}gico (CNPq-Brazil).   GNF
thanks CNPq for a PhD scholarship, and the Laboratoire Charles Coulomb
in Montpellier for the hospitality.  ROR also acknowledges support
from Funda\c{c}\~ao Carlos Chagas Filho de Amparo \`a Pesquisa do
Estado do Rio de Janeiro (FAPERJ), under grant No. E - 26 /
201.424/2014 and Coordena\c{c}\~ao de Pessoal de N\'{\i}vel Superior - CAPES
(Processo No. 88881.119017/2016-01).


\end{document}